\documentclass{amsart}
\def\Defined{}
\ifx\UseRussian\Defined
	\usepackage[T2A,T2B]{fontenc}
	\usepackage[cp1251]{inputenc}
	\usepackage[english,russian]{babel}
\fi
\usepackage{footmisc}
\usepackage[all]{xy}
\usepackage{color}
\definecolor{UrlColor}{rgb}{.9,0,.3}
\definecolor{SymbColor}{rgb}{.4,0,.9}
\definecolor{IndexColor}{rgb}{1,.3,.6}
\definecolor{eml1}{rgb}{.8,.1,.1}
\definecolor{eml2}{rgb}{.1,.6,.6}

\usepackage{xr-hyper}
\usepackage[unicode]{hyperref}
\hypersetup{pdfdisplaydoctitle=true}
\hypersetup{colorlinks}
\hypersetup{citecolor=UrlColor}
\hypersetup{urlcolor=UrlColor}
\hypersetup{pdffitwindow=true}
\hypersetup{pdfnewwindow=true}
\hypersetup{pdfstartview={FitH}}

\def\hyph{\penalty0\hskip0pt\relax-\penalty0\hskip0pt\relax}
\def\Hyph{-\penalty0\hskip0pt\relax}

\newcommand{\Basis}[1]{\overline{\overline{#1}}{}}
\newcommand{\Vector}[1]{\overline{#1}{}}
\newcommand{\gi}[1]{\boldsymbol{\textcolor{IndexColor}{#1}}}
\makeatletter
\newcommand{\NameDef}[1]{%
	\expandafter\gdef\csname #1\endcsname%
}%
\newcommand{\ShowSymbol}[1]{%
	\@nameuse{ViewSymbol#1}%
}%
\newcommand{\symb}[3]{%
	\@ifundefined{ViewSymbol#3}{%
		\NameDef{ViewSymbol#3}{\textcolor{SymbColor}{#1}}%
		\NameDef{RefSymbol#3}{\pageref{symbol: #3}}%
		\@namedef{LabelSymbol#3}{\label{symbol: #3}}%
	}{%
		\NameDef{RefSymbol#3}{}%
		\@namedef{LabelSymbol#3}{}%
	}%
	\ifcase#2
	\or
		$\@nameuse{ViewSymbol#3}$%
	\or
		\[\@nameuse{ViewSymbol#3}\]%
	\else%
	\fi%
	\@nameuse{LabelSymbol#3}%
}%
\makeatother

\newcommand{\subs}{${}_*$\Hyph}
\newcommand{\sups}{${}^*$\Hyph}

\newcommand{\CRstar}{{}_*{}^*}
\newcommand{\RCstar}{{}^*{}_*}

\newcommand{\RC}{$\RCstar$\Hyph}
\newcommand{\CR}{$\CRstar$\Hyph}
\newcommand{\drc}{$D\RCstar$\Hyph}
\newcommand{\dcr}{$D\CRstar$\hyph}
\newcommand{\rcd}{$\RCstar D$\Hyph}
\newcommand{\crd}{$\CRstar D$\Hyph}

\newcommand\sT{$\star T$\Hyph}%
\newcommand\Ts{$T\star$\Hyph}%

\renewcommand{\uppercasenonmath}[1]{}

\makeatletter
\newcommand\@dotsep{4.5}
\def\@tocline#1#2#3#4#5#6#7
{\relax
		\par \addpenalty\@secpenalty\addvspace{#2}%
		\begingroup \hyphenpenalty\@M
		\@ifempty{#4}{%
			\@tempdima\csname r@tocindent\number#1\endcsname\relax
		}{%
			\@tempdima#4\relax
		}%
		\parindent\z@ \leftskip#3\relax \advance\leftskip\@tempdima\relax
		\rightskip\@pnumwidth plus1em \parfillskip-\@pnumwidth
		#5\leavevmode\hskip-\@tempdima #6\relax
		\leaders\hbox{$\m@th
		\mkern \@dotsep mu\hbox{.}\mkern \@dotsep mu$}\hfill
		\hbox to\@pnumwidth{\@tocpagenum{#7}}\par
		\nobreak
		\endgroup
}
\makeatother 

\ifx\PrintBook\undefined
	\usepackage{fancyhdr}
	\makeatletter
	\renewcommand{\@indextitlestyle}{%
		\twocolumn[\section{\indexname}]%
		\def\IndexSpace{off}%
	}
	\makeatother 
	\thanks{\href{mailto:Aleks\_Kleyn@MailAPS.org}{Aleks\_Kleyn@MailAPS.org}}
\else
	\makeatletter
	\renewcommand{\@indextitlestyle}{%
		\twocolumn[\chapter{\indexname}]%
		\def\IndexSpace{off}%
		\let\@secnumber\@empty
		\chaptermark{\indexname}%
	}
	\makeatother 
	\email{\href{mailto:Aleks\_Kleyn@MailAPS.org}{Aleks\_Kleyn@MailAPS.org}}
\fi

\ifx\SelectlEnglish\undefined
	\ifx\UseRussian\undefined
		\def\SelectlEnglish{}
	\fi
\fi

\ifx\SelectlEnglish\undefined
	\newcommand\CurrentLanguage{Russian.}%
	\author{Александр Клейн}
	\newtheorem{theorem}{Теорема}[section]
	\newtheorem{corollary}[theorem]{Следствие}
	\theoremstyle{definition}
	\newtheorem{definition}[theorem]{Определение}
	\newtheorem{example}[theorem]{Пример}
	\newtheorem{xca}[theorem]{Exercise}
	\theoremstyle{remark}
	\newtheorem{remark}[theorem]{Замечание}
	
	\newcommand\Gbasis{$G$\Hyph базис}
	\newcommand\Gcoords{$G$\Hyph координат}
	\newcommand\Gspace{$G$\Hyph пространств}
	\newcommand\xRefDef[2]
		{
		\externaldocument[#1-Russian-]{#1.Russian}[http://arxiv.org/PS_cache/#2.pdf]
		\NameDef{xRefDef#1}{}%
		}
	\makeatletter
	\newcommand\xRef[2]%
	{%
		\@ifundefined{xRefDef#1}{%
		\ref{#2}%
		}{%
		\citeBib{#1}-\ref{#1-Russian-#2}%
		}%
	}%
	\newcommand\xEqRef[2]%
	{%
		\@ifundefined{xRefDef#1}{%
		\eqref{#2}%
		}{%
		\citeBib{#1}-\eqref{#1-Russian-#2}%
		}%
	}%
	\makeatother
	\ifx\PrintBook\undefined
		\newcommand{\BibTitle}{%
			\section{Список литературы}%
		}
	\else
		\newcommand{\BibTitle}{%
			\chapter{Список литературы}%
		}
	\fi
\else
	\newcommand\CurrentLanguage{English.}%
	\author{Aleks Kleyn}
	\newtheorem{theorem}{Theorem}[section]
	
	\theoremstyle{definition}
	\newtheorem{definition}[theorem]{Definition}

	\theoremstyle{remark}
	\newtheorem{remark}[theorem]{Remark}
	\newcommand\Gbasis{$G$\Hyph basis}
	\newcommand\Gcoords{$G$\Hyph coordinates}
	\newcommand\Gspace{$G$\Hyph space}
	\newcommand\xRefDef[2]
		{
		\externaldocument[#1-English-]{#1.English}[http://arxiv.org/PS_cache/#2.pdf]
		\NameDef{xRefDef#1}{}%
		}
	\makeatletter
	\newcommand\xRef[2]%
	{%
		\@ifundefined{xRefDef#1}{%
		\ref{#2}%
		}{%
		\citeBib{#1}-\ref{#1-English-#2}%
		}%
	}%
	\newcommand\xEqRef[2]%
	{%
		\@ifundefined{xRefDef#1}{%
		\eqref{#2}%
		}{%
		\citeBib{#1}-\eqref{#1-English-#2}%
		}%
	}%
	\makeatother
	\ifx\PrintBook\undefined
		\newcommand{\BibTitle}{%
			\section{References}%
		}
	\else
		\newcommand{\BibTitle}{%
			\chapter{References}%
		}
	\fi
\fi

\ifx\PrintBook\undefined
\else
	\numberwithin{section}{chapter}
\fi

\numberwithin{equation}{section}
\numberwithin{figure}{section}
\numberwithin{table}{section}
\numberwithin{Item}{section}
\numberwithin{Hfootnote}{section}

\makeatletter
\newcommand\org@maketitle{}
\let\org@maketitle\maketitle
\def\maketitle{%
	\hypersetup{pdftitle={\@title}}%
	\hypersetup{pdfauthor={\authors}}%
	\hypersetup{pdfsubject=\@keywords}%
	\org@maketitle
}
\def\make@stripped@name#1{%
	\begingroup
		\escapechar\m@ne
		\global\let\newname\@empty
		\protected@edef\Hy@tempa{\CurrentLanguage #1}%
		\edef\@tempb{%
			\noexpand\@tfor\noexpand\Hy@tempa:=%
			\expandafter\strip@prefix\meaning\Hy@tempa
		}%
		\@tempb\do{%
			\if\Hy@tempa\else
				\if\Hy@tempa\else
					\xdef\newname{\newname\Hy@tempa}%
				\fi
			\fi
		}%
	\endgroup
}%
\newenvironment{enumBib}{%
	\BibTitle
	\advance\@enumdepth \@ne
	\edef\@enumctr{enum\romannumeral\the\@enumdepth}\list
	{\csname biblabel\@enumctr\endcsname}{\usecounter
	{\@enumctr}\def\makelabel##1{\hss\llap{\upshape##1}}}
}{%
	\endlist
}

\def\Chapters#1{\ChapterList#1,LastChapter,}%
\def\LastChapter{LastChapter}%
\def\ChapterList#1,{\def\temp{#1}%
	\ifx\temp\LastChapter
	\else
		\@ifundefined{#1}{%
		}{%
			\def\Semafor{on}
		}
		\expandafter\ChapterList
	\fi
}%
\newcommand{\BiblioItem}[3]
{
	\def\Semafor{off}
	\Chapters{#1}
	\ifx\Semafor\ValueOn
		\ifx\IndexState\ValueOff
			\begin{enumBib}
			\def\IndexState{on}
		\fi
		\item \label{bibitem: #2}#3%
	\fi
}
\newcommand{\OpenBiblio}
{
	\def\IndexState{off}
}
\newcommand{\CloseBiblio}
{
	\ifx\IndexState\ValueOn
		\end{enumBib}
		\def\IndexState{off}
	\fi
}

\def\StartCite{[}%
\def\citeBib#1{\showCiteBib#1,endCite,}%
\def\endCite{endCite}%
\def\showCiteBib#1,{\def\temp{#1}%
\ifx\temp\endCite
]%
\def\StartCite{[}%
\else
	\StartCite\ref{bibitem: #1}%
	\def\StartCite{, }%
\expandafter\showCiteBib%
\fi}%

\makeatother 
\newcommand{\arp}{\ar @{-->}}

\newcommand{\bundle}[4]%
{%
	\def\tempa{}%
	\def\tempb{#3}%
	\def\tempc{#1}%
	\ifx\tempa\tempb%
		\ifx\tempa\tempc%
			#2%
		\else%
			\xymatrix{#2:#1\arp[r]&#4}%
		\fi%
	\else%
		\ifx\tempa\tempc%
			#2[#3]%
		\else%
			\xymatrix{#2[#3]:#1\arp[r]&#4}%
		\fi%
	\fi%
}%
\newcommand{\AddIndex}[2]%
{%
	{\bf #1}%
	\label{index: #2}%
}%
\newcommand{\Index}[3]%
{%
	\def\Semafor{off}%
	\Chapters{#1}%
	\ifx\Semafor\ValueOn%
		\def\tempa{}%
		\def\tempb{#3}%
		\ifx\IndexState\ValueOff%
			\begin{theindex}%
			\def\IndexState{on}%
		\fi%
		\ifx\IndexSpace\ValueOn%
			\indexspace%
			\def\IndexSpace{off}%
		\fi%
		\item #2%
		\ifx\tempa\tempb%
		\else%
			\ \pageref{index: #3}%
		\fi%
	\fi%
}%
\newcommand{\SubIndex}[3]
{
	\def\Semafor{off}
	\Chapters{#1}
	\ifx\Semafor\ValueOn
		\subitem #2 \pageref{index: #3}
	\fi
}%

\makeatletter
\newcommand{\Symb}[3]
{
	\def\Semafor{off}
	\Chapters{#1}
	\ifx\Semafor\ValueOn
		\ifx\IndexState\ValueOff
			\begin{theindex}
			\def\IndexState{on}
		\fi
		\ifx\IndexSpace\ValueOn
			\indexspace
			\def\IndexSpace{off}
		\fi
		\item $\@nameuse{ViewSymbol#3}$\ \ #2
		\@nameuse{RefSymbol#3}%
	\fi
}

\makeatother

\newcommand{\SetIndexSpace}%
{%
	\def\IndexSpace{on}%
}%
\def\ValueOff{off}
\def\ValueOn{on}

\newcommand{\OpenIndex}
{
	\def\IndexState{off}
}
\newcommand{\CloseIndex}
{
	\ifx\IndexState\ValueOn
		\end{theindex}
		\def\IndexState{off}
	\fi
}

\def\LastMemo{LastMemo}%
\def\MemoList#1//{\def\temp{#1}%
	\ifx\temp\LastMemo
	\else%
		\par\setlength{\parindent}{12pt}\textcolor{blue}{#1}%
		\expandafter\MemoList%
	\fi%
}%

\def\texPrefaceRefernceFrame{}
\xRefDef{0412.391}{math/pdf/0412/0412391v4}

\begin{document}

\title{Reference Frame in General Relativity}

\pdfbookmark[1]{Reference Frame in General Relativity}{TitleEnglish}
\begin{abstract}
A reference frame in event space is a smooth field of orthonormal bases.
Every reference frame is equipped by anholonomic coordinates.
Using anholonomic coordinates allows to find out relative speed
of two observers and appropriate Lorentz transformation.

Synchronization of a reference frame is an anholonomic time coordinate.
Simple calculations show how synchronization influences time measurement in
the vicinity of the Earth.

Measurement of Doppler shift from the star orbiting the black hole helps
to determine mass of the black hole. According observations of Sgr A,
if non orbiting observer estimates age of S2 about 10 Myr,
this star is 0.297 Myr younger.

\end{abstract}

\maketitle

\def\texIntro{}

\ifx\PrintBook\Defined
				\chapter{Introduction}
			\section{About This Book}
			\label{section: About This Book}

Sometimes is very hard to give name for book that you want to write.
Even you wrote it the whole life. And may be for this reason.
The way is not finished. However I feel it is time to share with people
my discoveries. Only future will tell which parts of this research
will be useful.
This book starts from learning of geometric object, turns to
reference frame in physics, then suddenly changes its course
to learn metric affine manifolds.

Since Einstein introduced general relativity, the close relation between geometry
and physics became a reality. At the same time, quantum mechanics introduced new concepts that
contradicted a tradition established during centuries.
All together, this meant that we need new geometrical concepts that will become part
of the language of quantum mechanics.
This is the reason for returning to the beginning.

The whole my life was dedicated to solve one of the greatest
mysteries that I met in the beginning of my life. When I learned
general relativity and quantum mechanics I felt that language of
quantum mechanics is not adequate to phenomenon that it observes.
I mean the geometry.

Я посвятил главу \ref{chapter: Space and Time in Physics}
to small essay that I wrote when I was young.\footnote{Unfortunately
some references are lost. If somebody recognize
familiar text, I will appreciate if he let me know exactly reference}
\fi

\ifx\texPrefaceTorsion\Defined
\section{Torsion}

Close relationship between the metric tensor and the connection is
the basis of the Riemann geometry.
В то же время, связность и метрика, как любой геометрический объект, являются объектом измерения.
When we study the general relativity equation, we introduce the lagrangian where
the metric tensor and the connection are independent. Later on, we discover that the connection
is symmetric and find dependence between connection and metric tensor. One of the
reasons is in the simplicity of the lagrangian.

\ifx\PrintBook\Defined
Глава \ref{chapter: General Relativistic Field Equations}
формально завершает книгу, хотя исторически она отображает
исследование, с которого начался определённый этап моей жизни.
В основу уравнений, выведенных в этой главе, было положено
непосредственное объединение двух лагранжианов: гравитационного
и квантового полей.
Повидимому столь прямолинейный подход
послужил причиной неудачи. Любая попытка решить уравнения
\eqref{eq: Field}, \eqref{eq: Einstein},
\eqref{eq: Cartan} и
\eqref{eq: Maxwell} приводит к тому,
что эта системв распадается на уравнение Эйнштейна и
уравнение поля, которые независимы.

Последующее развитие физики показало, что объединение общей теории
относительности и квантовой механики требует более мощных методов.
Основные направления на сегодняшний день можно разделить на теорию
струн и петлевую теорию гравитации.

Однако несмотря на неудачу, анализ полученных уравнений очень важен.
\fi

Analysis of quantum field theory shows that either
symmetry of connection or dependence between connection and metric may be broken. 
because неизбежны ошибки измерения.
As we expected this assumption leads particularly to space with torsion and nonzero covariant derivative
of metric.
We call the affine connected manifold with metric tensor
the metric-affine manifold \citeBib{Mielke}.

The metric-affine manifold appears in different physical applications.
Очень важно понять какова геометрия этого пространства,
как кручение может влиять на физические процессы.
Именно поэтому небольшая группа физиков продолжает изучать теорию
гравитации с кручением
\citeBib{Mielke,Obukhov,Sardanashvily,Gauge,Neeman}.

Independence of the metric tensor and the connection allows us to see which object is responsible for different
phenomenon in geometry and therefore in physics.

In particular we have two different definitions of a geodesic curve in the Riemann manifold.
Мы можем рассматривать геодезическую либо как кривую экстремальной длины
(соответствующую кривую мы называем extreme),
либо как кривую, вдоль которой касательный вектор переносится параллельно, оставаясь
касательным к кривой
(соответствующую кривую мы называем auto parallel).
Nonmetricity means that parallel transport does not conserve a length of vector and an angle between
vectors. This leads to a difference between
definitions of auto parallel and extreme lines (\citeBib{torsion} and
section \ref{section: Line with extreme length})
and to a change in the expression of the Frenet transport.
The change of geometry influences the second Newton law which we study in section \ref{NewtonLaw}.
I show in theorems \ref{theorem: LengthTangentVector} and \ref{theorem: First Newton law} 
that a free falling particle chooses an extreme line transporting
its momentum along the trajectory without change.

Форма второго закона Ньютона зависит от выбора формы потенциала.
В случае скалярного потенциала
the Newton second law holds the relationship between force, mass and acceleration.
В случае векторного потенциала
analysis of motion in a gravitational field shows that the field-strength tensor
depends on the derivative of the metric tensor.

Nonmetricity dramatically changes law how orthogonal frame moves in space.
However learning of parallel transport in space with nonmetricity allows us to
introduce the Cartan transport and an introduction of the connection compatible with the metric tensor
(section \ref{section: Cartan Transport}). The Cartan transport
holds the basis orthonormal and this makes it valuable tool
in dynamics (section \ref{NewtonLaw})
because the observer uses an orthonormal basis as his measurement tool.
The dynamics of a particle follows to the Cartan transport.
The question arises from this conclusion.
We can change the connection as we show above.
Why we need to learn manifolds with an arbitrary connection and the metric tensor?
First of all, the learning of the metric-affine manifold shows why everything
works well in the Riemann manifold and what changes in a general case.
Also physical constraints that appear in a model may lead to appearance of a nonmetricity
\citeBib{Gauge,Megged,gr-qc-9604027}.
Because the Cartan transport is the natural mechanism to conserve orthogonality
we expect that we will interpret a deviation of the test particle from the extreme line as
a result of an external to this particle
force\footnote{For instance if we extend the definition \eqref{eq: potential force} of a force
to a general case \eqref{eq: Newton} we can interpret a deviation of a charged particle
in electromagnetic field as result of the force
\[\ F^j=\frac e {cu^0} g^{ij} F_{li} u^l\]
The same way we can interpret a deviation of the auto parallel line as the force
\[F^i=-\frac {m c} {u^0} \Gamma(C)^i_{kl}u^ku^l\]
I remind that the Cartan symbol is the tensor}.
In this case the difference between two types of a transport
becomes measurable and meaningful. Otherwise another type
of a transport and nonmetricity are not observable and we can use
only the transport compatible with metric.

I see here one more opportunity. As follows from the paper \citeBib{Megged}
the torsion may depend on quantum properties of matter.
However the torsion is the part of the connection. This means that the connection may
also depend on quantum properties of matter. This may lead to breaking of
the Cartan transport. However this opportunity demands additional research.

The effects of torsion and a nonmetricity are cumulative.
They may be small but measurable. We can observe their effects not only in strong
fields like black hole or Big Bang but in regular conditions
as well.
Studying geometry and dynamics of point particle gives us a way to test this point of view.
There is mind to test this theory in condition when spin of quantum field is accumulated.
We can test a deviation from second Newton law or measure torsion by observing
the movement of 2 different particles.

To test if the spacetime has the torsion we
can test the opportunity to build a parallelogram in spacetime. 
We can conduct such test in strong electromagnetic field.
We can get 2 particles or 2 photons that start their movement from the same point
and using a mirror to force them to move along opposite sides of the parallelogram.
We can start this test when we do not have electromagnetic field and then repeat it in its presence.
If particles meet in the same place or we have the same interference then we have
torsion equal $0$ in this thread.
In particular, the torsion may influence the behavior of virtual particles.
\fi

\ifx\texPrefaceRefernceFrame\Defined
\section{Geometrical Object and Invariance Principle}

\ifx\PrintBook\Defined
Sections \ref{section: Basis in Vector Space}
and \ref{section: Geometrical Object of Vector Space}
was written under the great influence of the book \citeBib{Rashevsky}.
The studying of a homogenous space of a group of symmetry of a vector space
leads us to the definition of a basis of this space. A basis  manifold is a set of
bases of particular vector space and is an example of a homogenous space.
As it is shown in \citeBib{Rashevsky} it gives ability to define
concepts of invariance and of geometrical object.

We introduce two types of transformation of a basis manifold:
active and passive transformations. The difference between them is
that the passive transformation can be expressed as a transformation of
an original space.

This definition can be extended on an arbitrary manifold. However in
this case we generalize the definition of a basis and introduce a reference
frame. In case of an event space of general relativity it leads us to
a natural definition of a reference frame and the Lorentz transformation.
A reference frame in event space is a smooth field of orthonormal frames.
\fi

The invariance principle which we studied \citeBib{0412.391}
is limited by vector spaces and we can use it only
in frame of the special relativity. Our task is
to describe structures which allow to extend
the invariance principle to general relativity.

A measurement of a spatial interval and a time length is one of the major tasks of general relativity.
This is a physical process that allows the study of geometry in a certain area of spacetime.
From a geometrical point of view,
the observer uses an orthonormal basis in a tangent plane as his measurement tool
because an orthonormal basis leads to the simplest local geometry.
When the observer moves from point to point he brings his measurement tool with him.

Notion of a geometrical object is closely related with
physical values measured in space time.
The invariance principle allows expressing physical laws
independently from the selection of a basis.
On the other hand if we want
to examine a relationship obtained in the test, we need to select
measurement tool. In our case this is basis. Choosing
the basis we can define coordinates of the geometrical object
corresponding to studied physical value.
Hence we can define the measured value.

Every reference frame is equipped by anholonomic coordinates.
For instance, synchronization of a reference frame is an anholonomic time coordinate.
Simple calculations show how synchronization influences time measurement in
the vicinity of the Earth.
Measurement of Doppler shift from the star orbiting the black hole helps
to determine mass of the black hole.

Sections \ref{section: Time Delay in Central Body Gravitational Field}
and \ref{section: Lorentz Transformation in Orbital Direction}
show importance of calculations in orthogonal
frame. Coordinates that we use in event space are just labels
and calculations that we make in coordinates may appear
not reliable. For instance in papers \citeBib{Tartaglia,Tomozawa}
authors determine coordinate speed of light. This leads to
not reliable answer and as consequence of this to
the difference of speed of light in different directions.

Few authors consider Einstein's idea about variability of speed of light
\citeBib{Ranada}. However Einstein proposed this idea when supposed to create
theory of gravity in Minkovsky space, therefore he assumed that scale of
space and time does not change. When Einstein learned Riemann geometry
he changed his mind and never returned to idea about variable speed of
light. Scale of space and time and speed of light are correlated in present
theory and we cannot change one without changing another.

There are few papers dedicated to a variable speed of light theory
\citeBib{Magueijo,Bassett}. Their theory is based on idea that metric tensor
may be scaleable relative dilatation. This idea is not new.
As soon as Einstein published general relativity
Weyl introduced his idea to make theory invariant relative
conformal transformation. However Einstein was firmly opposed to
this idea because it broke dependence between distance
and proper time. You can find detailed analysis in \citeBib{Straumann}.

We have strong relation between the speed of light
and units of length and time in special and general relativity.
When we develop new theory and discover that speed of light is different
we should ask ourselves about the reason. Did we make accurate measurement?
Do we have alternative way to exchange information and synchronize reference
frame? Did transformations between reference frames change and do they create group?

In some models photon may have small rest mass \citeBib{Lammerzahl}.
In this case speed of light is different from maximal speed and may
depend on direction. Recent experiment \citeBib{Muller}
put limitation on parameters of these models.
\fi

\ifx\PrintBook\Defined
\section{Symmetry}

The torsion leads to a change in the Killing equation. We also need to add a similar equation
for the connection.

Ассиметрия связности возникает не только за счёт кручения. Изучение теории
геометрического объекта в векторном пространстве приводит к её естественному
обобщению: теории системы отсчёта в пространстве событий. Система отсчёта
является мощнейшим измерительным инструментом в общей теории относительности.
\fi

\def\texRefernceFrame{}

\ifx\PrintBook\Defined
				\chapter{Reference Frame in Event Space}
\fi

			\section{Reference Frame on Manifold}

As is shown in section
\xRef{0412.391}{section: Basis in Vector Space}
we can identify frame manifold of vector space
and symmetry group of this space.
Our interest  was not details of structure of basis,
and stated theory can be generalized
and extended on an arbitrary manifold.
The details of structure of basis did not interest us 
and stated theory can be generalized.
In this section we generalize the definition of a frame and introduce a reference
frame on a manifold.
In case of an event space of general relativity it leads us to
a natural definition of a reference frame and the Lorentz transformation.

When we study manifold $\mathcal{V}$ the geometry of tangent space
is one of important factors. 
In this section, we will make the following assumption.
\begin{itemize}
\item All tangent spaces have the same geometry.
\item Tangent space
is vector space $\mathcal{V}$ of finite dimension $n$.
\item Symmetry group of tangent space is Lie group $G$.
\end{itemize}

		\begin{definition}
		\label{definition: type G reference frame}
Set \symb{\Basis e=<\Vector e_{(i)},i\in I>}1{reference frame}
of vector fields \symb{\Vector e_{(i)}}1{vector field of reference frame}
is called \AddIndex{$G$\Hyph reference frame}{G reference frame}, if
for any $x\in\mathcal{V}$ set
$\Basis e(x)=<\Vector e_{(i)}(x),i\in I>$ is a $G$\Hyph basis\footnote{According to section
\xRef{0412.391}{section: Basis in Vector Space}
we can identify basis $\Basis e(x)$ with an element of group $G$.}
in tangent space $T_x$.\footnote{In each
particular case we need to prove existence of $G$\Hyph reference frame on manifold.}
We use notation
$\ShowSymbol{vector field of reference frame}\in\Basis e$
for vector fields which form
$G$\Hyph reference frame $\Basis e$.
		\qed
		\end{definition}

Vector field $\Vector a$ has expansion
	\begin{equation}
\Vector a=a^{(i)}\Vector e_{(i)}
	\label{eq: Vector field expansion, reference frame}
	\end{equation}
relative reference frame $\Basis e$.

If we do not limit definition of a reference frame by symmetry group,
then at each point of the manifold
we can select reference frame
\symb{\Basis\partial=<\Vector \partial_i>}1{coordinate reference frame}
based on vector fields tangent to lines $x^i=const$.
We call this field of bases
the \AddIndex{coordinate
reference frame}{coordinate reference frame}.
Vector field $\Vector a$ has expansion
	\begin{equation}
\Vector a=a^i\Vector \partial_i
	\label{eq: Vector field expansion, coordinate reference frame}
	\end{equation}
relative coordinate reference frame.
Then standard coordinates of reference frame $\Basis e$ have form
\symb{e^k_{(i)}}1{standard coordinates of reference frame}
	\begin{equation}
\Vector e_{(i)}=e^k_{(i)}\Vector \partial_k
	\label{eq: reference frame expansion relative coordinate reference frame}
	\end{equation}
Because vectors $\Vector e_{(i)}$ are linearly independent at each point matrix
$\|e^k_{(i)}\|$ has inverse matrix $\|e_k^{(i)}\|$
	\begin{equation}
\Vector \partial_k=e_k^{(i)}\Vector e_{(i)}
	\label{eq: coordinate reference frame expansion relative reference frame}
	\end{equation}

We use also a more extensive definition for
reference frame
on manifold, presented in form
\symb{\Basis e=(\Vector e_{(k)},\Vector e^{(k)})}1
{reference frame, extensive definition}
where we use the set of vector
fields $\Vector e_{(k)}$ and dual forms
\symb{\Vector e^{(k)}}1{dual forms, reference frame}
such that
	\begin{equation}
\Vector e^{(k)}(\Vector e_{(l)})=\delta^{(k)}_{(l)}
	\label{eq: dual forms and vector fields}
	\end{equation}
at each point.
Forms $\Vector e^{(k)}$ are defined uniquely from
\eqref{eq: dual forms and vector fields}.

In a similar way, we can introduce a
coordinate reference frame
\symb{(\Vector \partial_i,\Vector ds^i)}1
{coordinate reference frame, extensive definition}.
These reference frames are linked by the relationship
	\begin{align}
	\label{eq: holonomic and anholonomic vector field}
\Vector e_{(k)}&=e_{(k)}^i\Vector \partial_i\\
	\label{eq: holonomic and anholonomic forms}
\Vector e^{(k)}&=e^{(k)}_i\Vector dx^i
	\end{align}
From equations
\eqref{eq: holonomic and anholonomic vector field},
\eqref{eq: holonomic and anholonomic forms},
\eqref{eq: dual forms and vector fields} it follows
	\begin{equation}
e^{(k)}_ie^i_{(l)}=\delta^{(k)}_{(l)}
	\label{eq: frame and coframe}
	\end{equation}

In particular we assume that we have $GL(n)$-reference frame
$(\partial,dx)$ raised by $n$ differentiable vector fields
$\partial_i$ and 1-forms $dx^i$, that define field of bases $\partial$
and cobases $dx$ dual them.

If we have function $\varphi$ on $V$ than we define \AddIndex{pfaffian derivative}{pfaffian derivative}
\[d\varphi=\partial_i\varphi dx^i\]

\section{Reference Frame in Event Space}

Starting from this section, we consider orthogonal reference frame
$\Basis e=(\Vector e_{(k)},\Vector e^{(k)})$
in Riemann space with metric $g_{ij}$.
According to definition, at each point of Riemann space
vector fields of orthogonal reference frame satisfy to equation
\[
g_{ij}e^i_{(k)}e^j_{(l)}=g_{(k)(l)}
\]
where $g_{(k)(l)}=0$, if $(k)\ne(l)$, and $g_{(k)(k)}=1$
or $g_{(k)(k)}=-1$ depending on signature of metric.

We can define the
\AddIndex{reference frame in event space}
{reference frame in event space}
$V$ as $O(3,1)$\Hyph reference frame.
To enumerate vectors, we use index
$k=0, ..., 3$. Index $k=0$ corresponds to time like vector field.

		\begin{remark}
		\label{remark: existence of a reference frame}
We can prove the existence of a reference frame using the orthogonolization procedure at every point of space
time. From the same procedure we get that coordinates of basis smoothly depend on the point.

A smooth field of time like vectors of each basis defines congruence of lines that are tangent to this field.
We say that each line is a world line of an observer or a \AddIndex{local reference frame}{local reference frame}.
Therefore a reference frame is set of local reference frames.
		\qed
		\end{remark}

We define the \AddIndex{Lorentz transformation}{Lorentz transformation} as transformation of a reference frame
$$
{x'}^i = f^i (x^0, x^1, x^2, x^3)
$$
\begin{equation}
{e'}^i_{(k)} = a^i_j b^{(l)}_{(k)} e^j_{(l)}
\label{LorentzTransformation}
\end{equation}
where
$$
a^i_j = \frac {\partial x'^i} {\partial x'^j}
$$
$$
\delta_{(i)(l)} b^{(i)}_{(j)} b^{(l)}_{(k)} = \delta_{(j)(k)}
$$
We call the transformation $a^i_j$ the holonomic part and transformation $b^{(l)}_{(k)}$
the anholonomic part.

\section{Anholonomic Coordinates}

Let $E(V, G,\pi)$ be the principal bundle,
where $V$ is the differential manifold of dimension $n$ and class
not less than $2$. We also assume that $G$ is symmetry group of tangent plain.

We define connection form on principal bundle 
	\begin{equation}
\omega^L=\lambda^L_Nda^N+\Gamma^L_idx^i\ \ \ \omega=\lambda_Nda^N+\Gamma dx
	\label{eq: G connection form}
	\end{equation}
We call functions $\Gamma_i$ connection components.

If fiber is group $GL(n)$, than connection has form
	\begin{equation}
\omega^a_b=\Gamma^a_{bc}dx^c
	\label{eq: GLn connection form}
	\end{equation}
\[\Gamma^A_i=\Gamma^a_{bi}\]

A vector field $\Vector a$ has two types of coordinates: \AddIndex{holonomic coordinates}{vector holonomic coordinates}
\symb{a^i}1{vector holonomic coordinates} relative coordinate reference frame
and \AddIndex{anholonomic coordinates}{vector anholonomic coordinates}
\symb{a^{(i)}}1{vector anholonomic coordinates}
relative reference frame. These two forms of coordinates also hold the relation
	\begin{equation}
a^i(x)=e^i_{(i)}(x)a^{(i)}(x)
	\label{eq: parallel transfer, 3}
	\end{equation}
at any point $x$.

We can study parallel transfer of vector fields using any form of coordinates.
Because \eqref{LorentzTransformation} is a linear transformation we expect that parallel transfer in
anholonomic coordinates has the same representation as in holonomic coordinates.
Hence we write
\[
da^k=-\Gamma^k_{ij}a^idx^j
\]
\[
da^{(k)}=-\Gamma^{(k)}_{(i)(j)}a^{(i)}dx^{(j)}
\]
It is required to establish link between
\AddIndex{holonomic coordinate of connection}{holonomic coordinates of connection}
\symb{\Gamma^k_{ij}}1{holonomic coordinates of connection}
and \AddIndex{anholonomic coordinates of connection}{anholonomic coordinates of connection}
\symb{\Gamma^{(k)}_{(i)(j)}}1{anholonomic coordinates of connection}
	\begin{equation}
a^i(x+dx)=a^i(x)+da^i=c^i(x)-\Gamma^i_{kp}c^k(x)dx^p
	\label{eq: parallel transfer, 1}
	\end{equation}
	\begin{equation}
a^{(i)}(x+dx)=a^{(i)}(x)+da^{(i)}=c^{(i)}(x)-\Gamma^{(i)}_{(k)(p)}c^{(k)}(x)dx^{(p)}
	\label{eq: parallel transfer, 2}
	\end{equation}
Considering \eqref{eq: parallel transfer, 1},
\eqref{eq: parallel transfer, 2}, and \eqref{eq: parallel transfer, 3} we get
	\begin{equation}
\begin{matrix}
a^i(x)-\Gamma^i_{kp}a^k(x)dx^p\\
=e^i_{(i)}(x+dx)
\left(a^{(i)}(x)-\Gamma^{(i)}_{(k)(p)}e^{(k)}_i(x)a^i(x)e^{(p)}_p(x)dx^p\right)
\end{matrix}
	\label{eq: parallel transfer, 4}
	\end{equation}
It follows from \eqref{eq: parallel transfer, 4} that 
	\begin{align*}
\Gamma^{(i)}_{(k)(p)}e^{(k)}_i(x)e^{(p)}_p(x)a^i(x)dx^p
&=a^{(i)}(x)-e^{(i)}_i(x+dx)\left(a^i(x)-\Gamma^i_{kp}a^k(x)dx^p\right)\\
&=a^i(x)e^{(i)}_i(x)-e^{(i)}_i(x+dx)\left(a^i(x)-\Gamma^i_{kp}a^k(x)dx^p\right)\\
&=a^i(x)\left(e^{(i)}_i(x)-e^{(i)}_i(x+dx)\right)
+e^{(i)}_j(x)\Gamma^j_{ip}a^i(x)dx^p\\
&=e^{(i)}_j(x)\Gamma^j_{ip}a^i(x)dx^p
-a^i(x)\frac{\partial e^{(i)}_i(x)}{\partial x^p}dx^p\\
&=\left(e^{(i)}_j(x)\Gamma^j_{ip}
-\frac{\partial e^{(i)}_i(x)}{\partial x^p}\right)a^i(x)dx^p
	\end{align*}
Because $a^i(x)$ and $dx^p$ are arbitrary we get 
	\[
\Gamma^{(i)}_{(k)(p)}e^{(k)}_i(x)e^{(p)}_p(x)
=e^{(i)}_j(x)\Gamma^j_{ip}
-\frac{\partial e^{(i)}_i(x)}{\partial x^p}
	\]
	\begin{equation}
\Gamma^{(i)}_{(k)(p)}
=e^i_{(k)}e^p_{(p)}e^{(i)}_j\Gamma^j_{ip}
-e^i_{(k)}e^p_{(p)}\frac{\partial e^{(i)}_i}{\partial x^p}
	\label{eq: parallel transfer, 5}
	\end{equation}
We introduce symbolic operator
	\begin{equation}
\frac{\partial}{\partial x^{(p)}}
=e^p_{(p)}\frac{\partial}{\partial x^p}
	\label{eq: anholonomic derivative}
	\end{equation}
From \eqref{eq: frame and coframe} it follows
	\begin{equation}
e^i_{(l)}\frac{\partial e^{(k)}_i}{\partial x^p}+
e^{(k)}_i\frac{\partial e^i_{(l)}}{\partial x^p}=0
	\label{eq: parallel transfer, 6}
	\end{equation}
Substitude \eqref{eq: anholonomic derivative} and
\eqref{eq: parallel transfer, 6} into \eqref{eq: parallel transfer, 5}
\begin{equation}
\Gamma^{(i)}_{(k)(p)}=
e^i_{(k)}e^p_{(p)}e^{(i)}_j\Gamma^j_{ip}
-e^{(i)}_i\frac{\partial e^i_{(k)}}{\partial x^{(p)}}
\label{eq: Gamma1}
\end{equation}

Equation \eqref{eq: Gamma1} shows some similarity between holonomic and anholonomic coordinates.
We introduce symbol \symb{\partial_{(k)}}1{partial(k)} for the derivative along vector field $e_{(k)}$
$$\partial_{(k)}=e^i_{(k)}\partial_i$$
Then \eqref{eq: Gamma1} takes the form
$$
\Gamma^{(k)}_{(l)(p)}=
e^i_{(l)}e^r_{(p)}e^{(k)}_j\Gamma^j_{ir}
-e^i_{(l)}\partial_{(p)} e^{(k)}_i
$$

Therefore when we move from holonomic coordinates to anholonomic, the connection transforms
the way similarly to when we move from one coordinate system to another.
This leads us to the model of anholonomic coordinates.

The vector field $e_{(k)}$ generates lines defined by the differential equations
$$
e^j_{(l)} \frac{\partial t} {\partial x^j} = \delta^{(k)}_{(l)}
$$
or the symbolic system 
\begin{equation}
\frac{\partial t} {\partial x^{(l)}} = \delta^{(k)}_{(l)}
\label{DefForX}
\end{equation}
Keeping in mind the symbolic system \eqref{DefForX} we denote
the functional $t$ as \symb{x^{(k)}}1{x(k)}
and call it the
\AddIndex{anholonomic coordinate}{anholonomic coordinate}.
We call the regular coordinate holonomic.

From here we can find derivatives and get
\begin{equation}
\frac{\partial x^{(i)}} {\partial x^k} = e^{(i)}_k
\label{DerivativeForX}
\end{equation}
The necessary and sufficient conditions of complete integrability
of system \eqref{DerivativeForX} are
$$
c^{(i)}_{(k)(l)} =0
$$
where we introduced \AddIndex{anholonomity object}{anholonomity object}
\symb{c^{(i)}_{(k)(l)}}0{anholonomity object}
	\begin{equation}
\ShowSymbol{anholonomity object} =e^k_{(k)}e^l_{(l)}
\left(
\frac{\partial e^{(i)}_k}{\partial x^l} - \frac{\partial e^{(i)}_l}{\partial x^k}
\right)
	\label{eq: Anholonomity Object}
	\end{equation}
Therefore each reference frame has $n$ vector fields
$$
\partial_{(k)} = \frac\partial {\partial x^{(k)}} = e^i_{(k)} \partial_i
$$
which have commutator
$$
[\partial_{(i)},\partial_{(j)}]=
\left( e^k_{(i)}\partial_k e^l_{(j)} - e^k_{(j)}\partial_k e^l_{(i)}\right)e^{(m)}_l \partial_{(m)}=$$
$$e^k_{(i)}e^l_{(i)}\left( -\partial_k e_l^{(m)} + \partial_l e_l^{(m)}\right)\partial_{(m)}=
c^{(m)}_{(k)(l)}\partial_{(m)}
$$
For the same reason we introduce forms
$$dx^{(k)}=e^{(k)}=e^{(k)}_l dx^l$$
and an exterior differential of this form is
	\begin{equation}
	\begin{split}
d^2x^{(k)}&=d\left( e^{(k)}_i dx^i\right)\\
&=\left(\partial_j e_i^{(k)} -\partial_i e_j^{(k)}\right)dx^i \wedge dx^j\\
&=-c^{(m)}_{(k)(l)}dx^{(k)}\wedge dx^{(l)}
	\end{split}
	\label{eq: anholonomity}
	\end{equation}

Therefore when $c^{(i)}_{(k)(l)} \not =0$, the differential $dx^{(k)}$ is not an exact differential
and the system \eqref{DerivativeForX} in general cannot be integrated.
However we can create a meaningful object that
models the solution. We can study how function $x^{(i)}$ changes along different lines.
We call such cordinates \AddIndex{anholonomic coordinates on manifold}
{anholonomic coordinates on manifold}.

		\begin{remark}
		\label{remark: anholonomic coordinates, model}
Function $x^{(i)}$ is a natural parameter along a flow line of vector field $\Vector e_{(i)}$.
We study an instance of such function in section
\ref{section: Synchronization of Reference Frame}.
The proper time is defined along world line of local
reference frame. As we see in remark
\ref{remark: existence of a reference frame}
world lines of local
reference frames cover spacetime.
To make proper time of local
reference frames as time of reference frame
we expect that proper time smoothly changes
from point to point. To synchronize clocks of local
reference frames we use classical procedure of exchange light signals.

From mathematical point of view this is problem
to integrate differential form.
However, a change of function along a loop is
	\begin{equation}
	\begin{split}
\Delta x^{(i)}
&= \oint dx^{(i)}\\
&=\int\int c^{(i)}_{(k)(l)}dx^{(k)}\wedge dx^{(l)}\\
&=\int\int c^{(i)}_{(k)(l)} e^{(k)}_k e^{(l)}_l dx^k\wedge dx^l
	\end{split}
	\label{eq: change of coordinate along a loop}
	\end{equation}

Somebody may have impression that we cannot synchronize clock,
however this conflicts with our observation.
We accept that synchronization is possible until
we introduce time along non closed lines.
Synchronization breaks when we try synchronize clocks
along closed line.

This means ambiguity in definition of anholonomic coordinates.
		\qed
		\end{remark}

From now on we will not make a difference between holonomic and anholonomic coordinates.
Also, we will denote $b^{(l)}_{(k)}$ as ${a^{-1}}^{(l)}_{(k)}$ in the Lorentz transformation
\eqref{LorentzTransformation}.

Even form $dx^{(k)}$ is not exact differential, we can see
that form $d^2x^{(k)}$ is exterior differential of form $dx^{(k)}$.
Therefore
	\begin{equation}
	\label{eq: exterior derivative 3}
d^3x^{(k)}=0
	\end{equation}

We can represent exterior differential of form, written in anholonomic coordinates, as
\begin{align*}
&d(a_{(i_1)...(i_n)}dx^{(i_1)}\wedge ...\wedge dx^{(i_n)})\\
=&a_{(i_1)...(i_n),p}dx^p\wedge dx^{(i_1)}\wedge ...\wedge dx^{(i_n)}\\
-&a_{(i_1)...(i_n)}ddx^{(i_1)}\wedge ...\wedge dx^{(i_n)}-...
-(-1)^{n-1}a_{(i_1)...(i_n)}dx^{(i_1)}\wedge ...\wedge ddx^{(i_n)}\\
=&a_{(i_1)...(i_n),(p)}e_p^{(p)}e^p_{(r)}dx^{(r)}\wedge dx^{(i_1)}\wedge ...\wedge dx^{(i_n)}\\
-&a_{(i_1)...(i_n)}c^{(i_1)}_{(p)(r)}dx^{(p)}\wedge dx^{(r)}\wedge ...\wedge dx^{(i_n)}-...\\
-&(-1)^{n-1}a_{(i_1)...(i_n)}dx^{(i_1)}\wedge ...\wedge c^{(i_n)}_{(p)(r)}dx^{(p)}\wedge dx^{(r)}\\
=&(a_{(i_1)...(i_n),(p)}
-a_{(r)...(i_n)}c^{(r)}_{(p)(i_1)}-...
-a_{(i_1)...(r)} c^{(r)}_{(p)(i_n)})
dx^{(p)}\wedge dx^{(i_1)}\wedge ...\wedge dx^{(i_n)}
\end{align*}
In case of form $d^3x^{(k)}$ we get equation
\begin{equation}
	\label{eq: exterior derivative, anholonomic form}
\begin{array}{l}
d(c^{(k)}_{(i)(j)}dx^{(i)}\wedge dx^{(j)})\\
=(c^{(k)}_{(i)(j),(p)}
-c^{(k)}_{(r)(j)}c^{(r)}_{(p)(i)}
-c^{(k)}_{(i)(r)} c^{(r)}_{(p)(j)})
dx^{(p)}\wedge dx^{(i)}\wedge dx^{(j)}  
\end{array}
\end{equation}
From equations \eqref{eq: exterior derivative 3}
and \eqref{eq: exterior derivative, anholonomic form}
it follows
\begin{equation}
	\label{eq: exterior derivative, anholonomic form, 1}
(c^{(k)}_{(i)(j),(p)}
-c^{(k)}_{(r)(j)}c^{(r)}_{(p)(i)}
-c^{(k)}_{(i)(r)} c^{(r)}_{(p)(j)})
dx^{(p)}\wedge dx^{(i)}\wedge dx^{(j)}=0  
\end{equation}
It is easy to see that
\begin{equation}
	\label{eq: exterior derivative, anholonomic form, 2}
\begin{array}{l}
(-c^{(k)}_{(r)(j)}c^{(r)}_{(p)(i)}
-c^{(k)}_{(i)(r)} c^{(r)}_{(p)(j)})
dx^{(i)}\wedge dx^{(j)}\\
=(-c^{(k)}_{(r)(j)}c^{(r)}_{(p)(i)}
+c^{(k)}_{(r)(i)} c^{(r)}_{(p)(j)})
dx^{(i)}\wedge dx^{(j)}\\
=-2c^{(k)}_{(r)(j)}c^{(r)}_{(p)(i)}
dx^{(i)}\wedge dx^{(j)}
\end{array}
\end{equation}
Substituting from \eqref{eq: exterior derivative, anholonomic form, 2}
into \eqref{eq: exterior derivative, anholonomic form, 1} gives
\begin{equation}
	\label{eq: exterior derivative, anholonomic form, 3}
(c^{(k)}_{(i)(j),(p)}
-2c^{(k)}_{(r)(j)}c^{(r)}_{(p)(i)})
dx^{(p)}\wedge dx^{(i)}\wedge dx^{(j)}=  0
\end{equation}
From \eqref{eq: exterior derivative, anholonomic form, 3}, it follows
\begin{equation}
	\label{eq: exterior derivative, anholonomic form, 4}
\begin{array}{l}
c^{(k)}_{(i)(j),(p)}+c^{(k)}_{(j)(p),(i)}+c^{(k)}_{(p)(i),(j)}\\
=2c^{(k)}_{(r)(j)}c^{(r)}_{(p)(i)}
+2c^{(k)}_{(r)(p)}c^{(r)}_{(i)(j)}
+2c^{(k)}_{(r)(i)}c^{(r)}_{(j)(p)}
\end{array}
\end{equation}

We define the curvature form for connection \eqref{eq: G connection form}
\[\Omega=d\omega+[\omega,\omega]\]
\[\Omega^D=d\omega^D+C^D_{AB}\omega^A\wedge \omega^B=R^D_{ij}dx^i\wedge dx^j\]
where we defined a curvature object
\[R^D_{ij}=\partial_i\Gamma^D_j-\partial_j\Gamma^D_i+C^D_{AB}\Gamma^A_i\Gamma^B_j+\Gamma^D_k c^k_{ij}\]

The curvature form for the connection \eqref{eq: GLn connection form} is
	\begin{equation}
\Omega^a_c = d \omega^a_c + \omega^a_b \wedge \omega^b_c
	\label{eq: Curvature}
	\end{equation}
where we defined a curvature object
	\begin{equation}
R^D_{ij}=R^a_{bij}=\partial_i\Gamma^a_{bj}-\partial_j\Gamma^a_{bi}
+\Gamma^a_{ci}\Gamma^c_{bj}-\Gamma^a_{cj}\Gamma^c_{bi}+\Gamma^a_{bk} c^k_{ij}
	\label{eq: GLn curvature}
	\end{equation}
We introduce Ricci tensor
\[R_{bj}=R^a_{baj}=\partial_a\Gamma^a_{bj}-\partial_j\Gamma^a_{ba}
+\Gamma^a_{ca}\Gamma^c_{bj}-\Gamma^a_{cj}\Gamma^c_{ba}+\Gamma^a_{bk} c^k_{aj}\]

\ifx\texFuture\Defined
			\section{Covariant exterior Derivative}

exterior derivative of form \eqref{eq: holonomic and anholonomic forms}
is different from 0. At first sight it looks unusual. However
form \eqref{eq: holonomic and anholonomic forms} is not
exact differential.

We define covariant exterior derivative of form $t = t_a \theta^a$ as
\[
Dt = t_{a;b} \theta^b \land \theta^a
+ t_a c^a_{bc} \theta^b \land \theta^c
\]

When we look at such equation we ask if $d^2=0$ still is legal.

			\section{Moving Basis}

We meet another interesting situation when we study moving of frame along
manifold. If we introduce derivative for each vector of frame as
$$
\frac {De^i_{(k)}} {dx^l}
= e^i_{(k),l} + \Gamma^i_{pl} e^p_{(k)}
$$
then we get transformation for moving frame
$$
\frac {De^i_{(k)}} {dx^l}
= A^{(p)}_{(k)l} e^i_{(p)}
$$
where values $A_{(p)(k)l} = g_{(p)(r)} A^{(r)}_{(k)l}$ satisfy to equation
\[
A_{(p)(k)l} + A_{(k)(p)l} = - g_{(p)(k);l}
\]
In particular,
\[A_{(k)(k)l}= - \frac 1 2 g_{(k)(k);l}\]
We observed the same type of transformation when we analyzed Frenet transfer.

Therefore, if $g_{pk;l} = 0$, then we get orthogonal transformation of frame.
However this is not so in case of nonzero derivative.
\fi

\def\texGeomObject{}
\ifx\PrintBook\Defined
				\chapter{Geometrical Object}
\fi

			\section{Metric-affine Manifold}

For connection \xEqRef{0405.027}{eq: GLn connection form}
we defined the \AddIndex{torsion form}{torsion form}
	\begin{equation}
T^a = d^2 x^a + \omega^a_b \wedge dx^b
	\label{eq: Torsion}
	\end{equation}
From \xEqRef{0405.027}{eq: GLn connection form} it follows
	\begin{equation}
\omega^a_b \wedge dx^b=(\Gamma^a_{bc}-\Gamma^a_{cb})dx^c\wedge dx^b
	\label{eq: wedge connection}
	\end{equation}
Putting \eqref{eq: wedge connection} and \xEqRef{0405.027}{eq: anholonomity} into \eqref{eq: Torsion}
we get
	\begin{equation}
T^a=T^a_{cb}dx^c\wedge dx^b =
-c^a_{cb}dx^c\wedge dx^b + (\Gamma^a_{bc}-\Gamma^a_{cb})dx^c\wedge dx^b
\label{eq: Torsion 1}
	\end{equation}
where we defined \AddIndex{torsion tensor}{torsion tensor}
	\begin{equation}
T^a_{cb} =
\Gamma^a_{bc}-\Gamma^a_{cb}-c^a_{cb}
\label{eq: Torsion coordinates}
	\end{equation}

Commutator of second derivatives has form
	\begin{equation}
u^\alpha_{;kl}-u^\alpha_{;lk}
=R^\alpha_{\beta lk}u^\beta
-T^p_{lk}u^\alpha_{;p}
	\label{eq: commutator second derivative}
	\end{equation}
From \eqref{eq: commutator second derivative} it follows that
	\begin{equation}
	\label{eq: commutator second derivative of vector}
\xi^a_{;cb}-\xi^a_{;bc}
=R^a_{d bc}\xi^d
-T^p_{bc}\xi^a_{;p}
	\end{equation}

In Rieman space we have metric tensor $g_{ij}$ and connection $\Gamma^k_{ij}$.
One of the features of the Rieman space is symetricity of connection and covariant derivative
of metric is 0. This creates close relation between metric and connection.
However the connection is not necessarily symmetric
and the covariant derivative of the metric tensor may be different from 0.
In latter case we introduce the \AddIndex{nonmetricity}{nonmetricity}
	\begin{equation}
Q^{ij}_k=g^{ij}_{;k}=g^{ij}_{,k}+\Gamma^i_{pk}g^{pj}+\Gamma^j_{pk}g^{ip}
	\label{eq: Nonmetricity}
	\end{equation}

Due to the fact that derivative of the metric tensor is not 0, we cannot raise or lower index
of a tensor under derivative as we do it in regular Riemann space.
Now this operation changes to next
\[a^i_{;k} = g^{ij} a_{j;k} + g^{ij}_{;k} a_j\]
This equation for the metric tensor gets the following form
\[g^{ab}_{;k} = - g^{ai} g^{bj} g_{ij;k}\]

		\begin{definition}
We call a manifold with a torsion
and a nonmetricity
the \AddIndex{metric-affine manifold}{metric-affine manifold} \citeBib{Mielke}.
		\qed  
		\end{definition}

If we study a submanifold $V_n$ of a manifold $V_{n+m}$, we see that the immersion creates
the connection $\Gamma^\alpha_{\beta\gamma}$ that relates to the connection in manifold as
\[
\Gamma^\alpha_{\beta\gamma} e^l_\alpha =
\Gamma^l_{mk} e^m_\beta e^k_\gamma + \frac {\partial e^l_\beta} {\partial u^\gamma}
\]
Therefore there is no smooth immersion
of a space with torsion into the Riemann space.

			\section{Geometrical Meaning of Torsion}
Suppose that $a$ and $b$ are non collinear vectors in a point $A$ (see figure \ref{fig:Torsion}).

\begin{tabular}{l|r}
\begin{minipage}{160pt}
We draw the geodesic $L_a$ through the point $A$
using the vector $a$ as a tangent vector to $L_a$ in the point $A$.
Let $\tau$ be the canonical parameter on $L_a$ and
\[\frac {dx^k}{d\tau}=a^k\]
We transfer the vector $b$ along the geodesic $L_a$ from the point $A$
into a point $B$ that defined by any value of the parameter $\tau=\rho>0$. We mark the result as $b'$.
\end{minipage}
&
\begin{minipage}{160pt}
We draw the geodesic $L_b$ through the point $A$
using the vector $b$ as a tangent vector to $L_b$ in the point $A$.
Let $\varphi$ be the canonical parameter on $L_b$ and
\[\frac {dx^k}{d\varphi}=b^k\]
We transfer the vector $a$ along the geodesic $L_b$ from the point $A$
into a point $D$ that defined by any value of the parameter $\varphi=\rho>0$. We mark the result as $a'$.
\end{minipage}
\end{tabular}

\begin{tabular}{l|r}
\begin{minipage}{160pt}
We draw the geodesic $L_{b'}$ through the point $B$
using the vector $b'$ as a tangent vector to $L_{b'}$ in the point $B$.
Let $\varphi'$ be the canonical parameter on $L_{b'}$ and 
\[\frac {dx^k}{d\varphi'}=b'^k\]
We define a point $C$ on the geodesic $L_{b'}$ by parameter value $\varphi'=\rho$
\end{minipage}
&
\begin{minipage}{160pt}
We draw the geodesic $L_{a'}$ through the point $D$
using the vector $a'$ as a tangent vector to $L_{a'}$ in the point $D$.
Let $\tau'$ be the canonical parameter on $L_{a'}$ and 
\[\frac {dx^k}{d\tau'}=a'^k\]
We define a point $E$ on the geodesic $L_{a'}$ by parameter value $\tau'=\rho$
\end{minipage}
\end{tabular}

Formally lines $AB$ and $DE$ as well as lines $AD$ and $BC$ are parallel
lines. Lengths of $AB$ and $DE$ are the same as well as lengths
of $AD$ and $BC$ are the same. We call this figure
a \AddIndex{parallelogram}{parallelogram} based on vectors
$a$ and $b$ with the origin in the point $A$.

		\begin{theorem}
Suppose $CBADE$ is a parallelogram with a origin in
the point $A$; then the resulting figure will not be closed \citeBib{torsion}.
The value of the difference of coordinates
of points $C$ and $E$ is equal to surface integral of the torsion
over this parallelogram\footnote{Proof of this statement I found in \citeBib{Shilov}}
\[\Delta_{CE}x^k=\iint T^k_{mn}dx^m \wedge dx^n \]
		\end{theorem}

	\begin{figure}
	\begin{center}
	\begin{picture}(150,150)
\put( 20, 25){ $A$ }
\put( 30, 65){ $a$ }
\put( 45, 50){ $b$ }
\put( 55, 130){ $B$ }
\put( 80, 135){ $b'$ }
\put( 155, 80){ $a'$ }
\put( 165, 150){ $C$ }
\put( 170, 40){ $D$ }
\put( 195, 135){ $E$ }
\put(-115, -65){\qbezier(150,100)(162,150)(185,190)}
\put(35, 35){\vector(1,4){11}}
\put(-115, -65){\qbezier(185,190)(218,207)(283,210)}
\put(70, 125){\vector(2,1){30}}
\put(169, 145){\line(3, -2){18}}
\put(-149, -154){\qbezier(320,212)(323,260)(336,287)}
\put(171, 59){\vector(0,1){35}}
\put(-149, -154){\qbezier(185,190)(208,206)(320,212)}
\put(35, 35){\vector(3,2){30}}
	\end{picture}
	\caption{Meaning of Torsion}
	\label{fig:Torsion}
	\end{center}
	\end{figure}
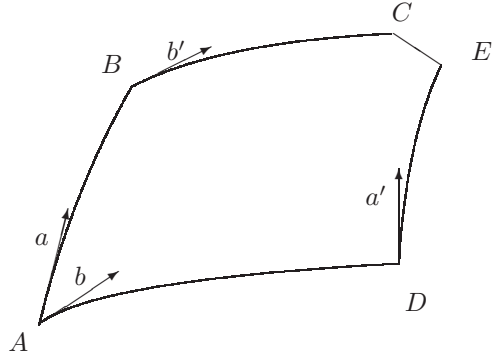
		\begin{proof}
We can find an increase of coordinate $x^k$ along any geodesic as
	\begin{align*}
\Delta x^k&=\frac{dx^k}{d\tau}\tau+\frac 1 2 \frac{d^2 x^k}{d\tau^2}\tau^2+O(\tau^2)=\\
&=\frac{dx^k}{d\tau}\tau-\frac 1 2 \Gamma^k_{mn}\frac{dx^m}{d\tau} \frac{dx^n}{d\tau}\tau^2+O(\tau^2)
	\end{align*}
where $\tau$ is canonical parameter and we take values of derivatives and components
$\Gamma^k_{mn}$ in the initial point. In particular we have 
	\[
\Delta_{AB} x^k=a^k\rho-\frac 1 2 \Gamma^k_{mn}(A)a^m a^n\rho^2+O(\rho^2)
	\]
along the geodesic $L_a$ and
	\begin{equation}
\Delta_{BC} x^k=b'^k\rho-\frac 1 2 \Gamma^k_{mn}(B)b'^m b'^n\rho^2+O(\rho^2)
	\label{eq:Torsion_1}
	\end{equation}
along the geodesic $L_{b'}$. Here
	\begin{equation}
b'^k=b^k- \Gamma^k_{mn}(A)b^m dx^n+O(dx)
	\label{eq:Torsion_2}
	\end{equation}
is the result of parallel transfer of $b^k$ from $A$ to $B$ and
	\begin{equation}
dx^k=\Delta_{AB} x^k=a^k\rho
	\label{eq:Torsion_3}
	\end{equation}
with precision of small value of first level. Putting
\eqref{eq:Torsion_3} into \eqref{eq:Torsion_2} and \eqref{eq:Torsion_2} into
\eqref{eq:Torsion_1} we will receive
\[
\Delta_{BC} x^k=b^k \rho- \Gamma^k_{mn}(A)b^m a^n \rho^2
-\frac 1 2 \Gamma^k_{mn}(B)b^m b^n\rho^2+O(\rho^2)
\]
Common increase of coordinate $x^K$ along the way $ABC$ has form
\[
\Delta_{ABC}x^k=\Delta_{AB} x^k+\Delta_{BC} x^k=
\]
	\begin{equation}
	\label{eq:Torsion_4}
=(a^k+b^k)\rho- \Gamma^k_{mn}(A)b^m a^n \rho^2-
	\end{equation}
\[
-\frac 1 2 \Gamma^k_{mn}(B)b^m b^n\rho^2
-\frac 1 2 \Gamma^k_{mn}(A)a^m a^n\rho^2+O(\rho^2)
\]

Similar way common increase of coordinate $x^K$ along the way $ADE$ has form
\[
\Delta_{ADE}x^k=\Delta_{AD} x^k+\Delta_{DE} x^k=
\]
	\begin{equation}
	\label{eq:Torsion_5}
=(a^k+b^k)\rho- \Gamma^k_{mn}(A)a^m b^n \rho^2-
	\end{equation}
\[
-\frac 1 2 \Gamma^k_{mn}(D)a^m a^n\rho^2
-\frac 1 2 \Gamma^k_{mn}(A)b^m b^n\rho^2+O(\rho^2)
\]

From \eqref{eq:Torsion_4} and \eqref{eq:Torsion_5}, it follows that
\[
\Delta_{ADE}x^k-\Delta_{ABC}x^k=\]
\[= \Gamma^k_{mn}(A)b^m a^n\rho^2
+\underline{\frac 1 2 \Gamma^k_{mn}(B)b^m b^n\rho^2}_1
+\underline{\frac 1 2 \Gamma^k_{mn}(A)a^m a^n\rho^2}_2-\]
\[- \Gamma^k_{mn}(A)a^m b^n \rho^2
-\underline{\frac 1 2 \Gamma^k_{mn}(D)a^m a^n\rho^2}_2
-\underline{\frac 1 2 \Gamma^k_{mn}(A)b^m b^n\rho^2}_1+O(\rho^2)
\]
For small enough value of $\rho$ underlined terms annihilate each other
and we get integral sum for expression
\[
\Delta_{ADE}x^k-\Delta_{ABC}x^k=\iint_\Sigma (\Gamma^k_{nm}- \Gamma^k_{mn})dx^m \wedge dx^n
\]

However it is not enough to find the difference
\[\Delta_{ADE}x^k-\Delta_{ABC}x^k\]
to find the difference of coordinates of points $C$ and $E$.
Coordinates may be anholonomic and we have to consider that
coordinates along closed loop change \xEqRef{0405.027}{eq: change of coordinate along a loop}
\[\Delta x^k=\oint_{ECBADE}dx^k=-\iint_\Sigma c^k_{mn}dx^m \wedge dx^n \]
where $c$ is anholonomity object.

Finally the difference of coordinates of points $C$ and $E$ is
\[\Delta_{CE}x^k=\Delta_{ADE}x^k-\Delta_{ABC}x^k+\Delta x^k=
\iint_\Sigma (\Gamma^k_{nm}- \Gamma^k_{mn}-c^k_{mn})dx^m \wedge dx^n\]
Using \eqref{eq: Torsion coordinates} we prove the statement. 
		\end{proof}

				\section{Relation between Connection and Metric}

Now we want to find how we can express connection if we know
metric and torsion. According to definition
\[-Q_{kij}=g_{ij;k}=g_{ij,k}-\Gamma^p_{ik}g_{pj}-\Gamma^p_{jk}g_{pi}\]
\[-Q_{kij}=g_{ij,k}-\Gamma^p_{ik}g_{pj}-\Gamma^p_{kj}g_{pi}-S^p_{jk}g_{pi}\]
We move derivative of $g$ and torsion to the left-hand side.
	\begin{equation}
g_{ij,k}+Q_{kij}-S^p_{jk}g_{pi}=\Gamma^p_{ik}g_{pj}+\Gamma^p_{kj}g_{pi}
	\label{eq: Connection1}
	\end{equation}
Changing order of indexes we write two more equations
	\begin{equation}
g_{jk,i}+Q_{ijk}-S^p_{ki}g_{pj}=\Gamma^p_{ji}g_{pk}+\Gamma^p_{ik}g_{pj}
	\label{eq: Connection2}
	\end{equation}
	\begin{equation}
g_{ki,j}+Q_{jki}-S^p_{ij}g_{pk}=\Gamma^p_{kj}g_{pi}+\Gamma^p_{ji}g_{pk}
	\label{eq: Connection3}
	\end{equation}
If we substruct equation \eqref{eq: Connection1} from sum of equations
\eqref{eq: Connection2} and \eqref{eq: Connection1} we get
\[
g_{ki,j}+g_{jk,i}-g_{ij,k}+Q_{ijk}+Q_{jki}-Q_{kij}
-S^p_{ij}g_{pk}-S^p_{ki}g_{pj}+S^p_{jk}g_{pi}=2\Gamma^p_{ji}g_{pk}
\]
Finally we get
\[
\Gamma^p_{ji}=\frac 1 2 g^{pk}
(g_{ki,j}+g_{jk,i}-g_{ij,k}+Q_{ijk}+Q_{jki}-Q_{kij}
-S^r_{ij}g_{rk}-S^r_{ki}g_{rj}+S^r_{jk}g_{ri})
\]

\def\texGenRelativity{}
\ifx\PrintBook\Defined
				\chapter{Application in General Relativity}
\fi

			\section{Synchronization of Reference Frame}
			\label{section: Synchronization of Reference Frame}

Because an observer uses an orthogonal basis for measurement at each point
we can expect that he uses anholonomic coordinates as well.
We also see that the time coordinate along a local reference frame is the observer's proper time.
Because the reference frame consists of local reference frames, we expect that their proper times
are synchronized.

We introduce the \AddIndex{synchronization of reference frame}{synchronization of reference frame}
as the anholonomic time coordinate.

Because synchronization is the anholonomic coordinate it introduces new physical phenomena
that we should keep in mind when working with strong gravitational fields or making
precise measurements. I describe one of these phenomena below.

		\section{Anholonomic Coordinates in Central Body Gravitational Field}
		\label{section: Anholonomic Coordinates in Central Body Gravitational Field}

We will study an observer orbiting around a central body.
The results are only estimation and are good when eccentricity is near $0$
because we study circular orbits.
However, the main goal of this estimation is to show that we have a measurable effect of anholonomity.

We use the Schwarzschild metric of a central body
	\begin{equation}
ds^2 = \frac{r-r_g} r c^2 dt^2
- \frac r {r-r_g} dr^2 - r^2 d\phi^2 - r^2 sin^2 \phi d\theta^2
	\label{eq: Schwarzschild}
	\end{equation}
$$r_g = \frac {2 G m} {c^2}$$
$G$ is the gravitational constant, $m$ is the mass of the central body,
$c$ is the speed of light.

Connection in this metric is
\[\Gamma^0_{10}=\frac {r_g} {2r(r-r_g)}\]
\[\Gamma^1_{00}=\frac {r_g(r-r_g)} {2r^3}\]
\[\Gamma^1_{11}=-\frac {r_g} {2r(r-r_g)}\]
\[\Gamma^1_{22}=-(r-r_g)\]
\[\Gamma^1_{33}=-(r-r_g)\sin^2\phi\]
\[\Gamma^2_{12}=-\frac 1 r\]
\[\Gamma^2_{33}=-\sin\phi\cos\phi\]
\[\Gamma^3_{13}=-\frac 1 r\]
\[\Gamma^3_{23}=\cot\phi\]


I want to show one more way
to calculate the Doppler shift. The Doppler shift in gravitational field is
well known issue, however the method that I show is useful to
better understand physics of gravitational field.

We can describe the movement of photon in gravitational field
using its wave vector $k^i$. The length of this vector is 0;
$\frac {k^i} {dx^i} = const$; a trajectory is
geodesic and therefore coordinates of this vector satisfy to differential equation
	\begin{equation}
dk^i = -\Gamma^i_{kl}k^k dx^l
	\label{eq: Transport}
	\end{equation}

We looking for the frequency $\omega$ of light
and $k^0$ is proportional $\omega$.
Let us consider the radial movement of a photon. In this case wave vector has form 
$k=(k^0,k^1,0,0)$.
In the central field with metric
\eqref{eq: Schwarzschild} we can choose
\[k^0=\frac \omega c \sqrt{\frac r {r-r_g}}\]
\[k^1=\omega \sqrt{\frac {r-r_g} r}\]
\[dt=\frac {k^0} {k^1} dr= \frac 1 c \frac r {r-r_g}dr\]
Then the equation \eqref{eq: Transport} gets form
\[dk^0=-\Gamma^0_{10}(k^1 dt + k^0 dr)\]
\[d\left(\frac\omega c \sqrt{\frac r {r-r_g}}\right)=
-\frac{r_g \omega} {2r(r-r_g)}\left(\sqrt{\frac {r-r_g} r}\frac r {r-r_g}
+\sqrt{\frac r {r-r_g}}\right)\frac {dr} c\]
\[d\omega \sqrt{\frac r {r-r_g}}-\omega\frac 1 2 \sqrt{\frac {r-r_g}r}
\frac {r_gdr} {(r-r_g)^2}
=-\frac {r_g\omega dr} {r(r-r_g)}\sqrt{\frac r {r-r_g}}\]
\[\frac  {d\omega} \omega
=-\frac {r_g} {2r(r-r_g)}dr\]
\[\ln  \omega
=\frac 1 2 \ln \frac r {r-r_g} + \ln C\]
If we define $\omega = \omega_0$ when $r=\infty$, we get finally 
\[\omega
=\omega_0\sqrt{\frac r {r-r_g}}\]

			\subsection{Time Delay in Central Body Gravitational Field}
			\label{section: Time Delay in Central Body Gravitational Field}

We will study orbiting around a central body.
The results are only an estimation and are good when eccentricity is near $0$
because we study circular orbits.
However, the main goal of this estimation is to show that we have a measurable effect of anholonomity.

Let us compare the measurements of two observers. The first observer fixed his position in the gravitational field
$$t=\frac s  c \sqrt{\frac r {r-r_g}}$$ $$r=const, \phi=const, \theta=const$$
The second observer orbits the center of the field with constant speed
$$t=s \sqrt{\frac r {(r-r_g) c^2 - \alpha^2 r^3}}$$
$$\phi=\alpha\ s \sqrt{\frac r {(r-r_g) c^2 - \alpha^2 r^3}}$$
$$r=const,\theta=const$$
We choose a natural parameter for both observers.

The second observer starts his travel when $s=0$ and finishes it when returning to the same spatial point.
Because $\phi$ is a cyclic coordinate the second observer finishes his travel when $\phi=2\pi$. We have at this point
$$s_2=\frac{2\pi} \alpha \sqrt{\frac{(r-r_g) c^2 - \alpha^2 r^3} r}$$
$$t=T=\frac{2\pi} \alpha$$
The value of the natural parameter for the first observer at this point is
$$s_1=\frac{2\pi} \alpha c\sqrt{\frac{r-r_g} r}$$
The difference between their proper times is
$$
\Delta s = s_1 - s_2
=\frac{2\pi} \alpha
\left(c\sqrt{\frac{r-r_g} r}-\sqrt{\frac{(r-r_g) c^2 - \alpha^2 r^3} r}\right)
$$
We have a difference in centimeters. To get this difference in seconds we should divide both sides by c.
$$\Delta t = \frac{2\pi} \alpha
\left(\sqrt{\frac{r-r_g} r}-\sqrt{\frac{r-r_g} r - \frac{\alpha^2 r^2} {c^2}}\right)$$

Now we get specific data.

The mass of the Sun is $1.989_{10}33$ g, the Earth orbits the Sun at a
distance of $1.495985_{10}13$ cm from its center during $365.257$ days. In this case we get
$\Delta t = 0.155750625445089$ s. Mercury orbits the Sun at a
distance of $5.791_{10}12$ cm from its center during $58.6462$ days. In this case we get
$\Delta t = 0.145358734930827$ s.

The mass of the Earth is $5.977_{10}27$ g. The spaceship that orbits the Earth at a
distance of $6.916_{10}8$ cm from its center during $95.6$ mins has $\Delta t = 1.8318_{10}-6$ s.
The Moon orbits the Earth at a
distance of $3.84_{10}10$ cm from its center during $27.32$ days. In this case we get
$\Delta t = 1.372_{10}-5$ s.

For better presentation I put these data to
tables \ref{table: Effect of Anholonomic Coordinates, Sun},
\ref{table: Effect of Anholonomic Coordinates, Earth},
and \ref{table: Effect of Anholonomic Coordinates, Sgr A}.

Because clocks of first observer show larger time at meeting,
first observer estimates age of second one
older then real. Hence, if we get parameters of S2 orbit from \citeBib{Ghez},
we get that if first observer estimates age of S2 as 10 Myr then
S2 will be .297 Myr younger.

\begin{table}[h]
\caption{Sun is central body, mass is $1.989_{10}33$ g}
\label{table: Effect of Anholonomic Coordinates, Sun}
\noindent\[
\begin{tabular}{|c|c|c|}
\hline
Sputnik&Earth&Mercury\\
\hline
distance, cm&$1.495985_{10}13$&$5.791_{10}12$\\
\hline
orbit period, days&$365.257$&$58.6462$\\
\hline
Time delay s&$0.15575$&$0.14536$\\
\hline
\end{tabular}
\]
\end{table}

\begin{table}[h]
\caption{Earth is central body, mass is $5.977_{10}27$ g}
\label{table: Effect of Anholonomic Coordinates, Earth}
\noindent\[
\begin{tabular}{|c|c|c|}
\hline
Sputnik&spaceship&Moon\\
\hline
distance, cm&$6.916_{10}8$&$3.84_{10}10$\\
\hline
orbit period&$95.6$ mins&$27.32$ days\\
\hline
Time delay, s&$1.8318_{10}-6$&$1.372_{10}-5$\\
\hline
\end{tabular}
\]
\end{table}

\begin{table}[h]
\caption{Sgr A is central body, S2 is sputnik}
\label{table: Effect of Anholonomic Coordinates, Sgr A}
\noindent\[
\begin{tabular}{|c|c|c|}
\hline
mass, $M_\odot$&$4.1_{10}6$&$3.7_{10}6$\\
\hline
distance sm&$1.4692_{10}16$&$1.1565_{10}16$\\
\hline
orbit period, years&$15.2$&$15.2$\\
\hline
Time delay, min&$164.7295$&$153.8326$\\
\hline
\end{tabular}
\]
\end{table}

			\section{Lorentz Transformation in Orbital Direction}
			\label{section: Lorentz Transformation in Orbital Direction}

The reason for the time delay that we estimated above is in Lorentz transformation between
stationary and orbiting observers.
This means that we have rotation in plain $(e_{(0)}, e_{(2)})$.
The basis vectors for stationary observer are
\[e_{(0)}=(\frac 1 c \sqrt{\frac r {r-a}}, 0, 0, 0)\]
\[e_{(2)}=(0, 0, \frac 1 r, 0)\]
We assume that for orbiting observer changes of $\phi$ and $t$
are proportional and
\[d\phi=\omega dt\]
Unit vector of speed in this case should be proportional to vector
	\begin{equation}
(1, 0, \omega, 0)
	\label{NewVectorE2}
	\end{equation}
The length of this vector is
	\begin{equation}
L^2=\frac {r-a} r c^2 - r^2 \omega^2
	\label{LengthNewVectorE2}
	\end{equation}
We see in this expression very familiar pattern and expect that linear
speed of orbiting observer is $V = \omega r$.

However we have to remember that we make measurement in gravitational
field and coordinates are just tags to label points in spacetime.
This means that we need a legal method to measure speed.

If an object moves from point $(t, \phi)$ to point $(t+dt, \phi+d\phi)$
we need to measure spatial and time intervals between these points.
We assume that in both points there are observers $A$ and $B$.
Observer $A$ sends the same time light signal to $B$ and ball that has
angular speed $\omega$. Whatever observer $B$ receives, he sends light signal
back to $A$.

When $A$ receives first signal he can estimate distance to $B$.
When $A$ receives second signal he can estimate how long ball moved to $B$.

The time of travel of light in both directions is the same.
Trajectory of light is determined by equation $ds^2=0$.
In our case we have
\[\frac {r-r_g} r c^2 dt^2 - r^2 d\phi^2 = 0\]
When light returns back to observer $A$ the change of $t$ is
\[dt=2 \sqrt{\frac r {r-r_g}} c^{-1}r d\phi\]
The proper time of first observer is
\[ds^2= \frac {r-r_g} r c^2 4 \frac r {r-r_g}c^{-2}r^2 d\phi^2\]
Therefore spatial distance is
\[L=r d\phi\]
When object moving with angular speed $\omega$ gets to $B$
change of $t$ is $\frac {d\phi} \omega$.
The proper time at this point is 
\[ds^2 = \frac {r - r_g} r c^2 d\phi^2 \omega^{-2}\]
\[T = \sqrt{\frac {r - r_g} r} \omega^{-1} d\phi\]
Therefore the observer $A$ measures speed
\[V=\frac L T = \sqrt{\frac r {r - r_g}} r\omega\]

We can use speed $V$ as parameter of Lorentz transformation. Then length
\eqref{LengthNewVectorE2} of vector \eqref{NewVectorE2} is
\[L=\sqrt{\frac {r-r_g} r \left(c^2 - \frac r {r-r_g}r^2 \omega^2\right)}=
\sqrt{\frac {r-r_g} r} c\sqrt{\left(1 - \frac {V^2}{c^2}\right)}\]
Therefore time ort of moving observer is
\[e'_{(0)}=\left(\frac 1 L, 0, \frac \omega L, 0\right)\]
\[e'_{(0)}=\left(\sqrt{\frac r {r-r_g}} c^{-1}\frac 1 {\sqrt{\left(1 - \frac {V^2}{c^2}\right)}}, 0,
\omega \sqrt{\frac r {r-r_g}} c^{-1}\frac 1 {\sqrt{\left(1 - \frac {V^2}{c^2}\right)}}, 0\right)\]
Spatial ort $e'_{(2)}=(A,0,B,0)$ is orthogonal $e'_{(0)}$ and has length $-1$. Therefore
	\begin{equation}
\frac {r-r_g} r c^2\frac 1 L A - r^2 \frac \omega L B = 0
	\label{NewVectorE0_1}
	\end{equation}
	\begin{equation}
\frac {r-r_g} r c^2 A^2 - r^2 B^2 = -1
	\label{NewVectorE0_2}
	\end{equation}
We can express $A$ from \eqref{NewVectorE0_1}
\[A = c^{-2}\frac r{r-r_g} r^2 \omega B\]
and substitute into \eqref{NewVectorE0_2}
\[ c^{-2}\frac r{r-r_g} r^4 \omega^2 B^2 - r^2 B^2 = -1\]
\[\frac {V^2}{c^2} r^2 B^2 - r^2 B^2 = -1\]
Finally spatial ort in direction of movement is
\[e'_{(2)}=\left(c^{-2}\frac r{r-r_g} r \omega \frac 1 {\sqrt{1 - \frac {V^2}{c^2}}},
0,\frac 1 r \frac 1 {\sqrt{1 - \frac {V^2}{c^2}}},0\right)\]
\[e'_{(2)}=\left(c^{-2}\sqrt{\frac r{r-r_g}} \frac V {\sqrt{1 - \frac {V^2}{c^2}}},
0,\frac 1 r \frac 1 {\sqrt{1 - \frac {V^2}{c^2}}},0\right)\]

Therefore we get transformation
	\begin{equation}
	\begin{split}
{e'}_{(0)}=\frac 1 {\sqrt{1 - \frac {V^2}{c^2}}} e_{(0)} +
\frac V  c \frac 1 {\sqrt{1 - \frac {V^2}{c^2}}} e_{(2)}
\\
{e'}_{(2)}=\frac V c \frac 1 {\sqrt{1 - \frac {V^2}{c^2}}} e_{(0)} +
\frac 1 {\sqrt{1 - \frac {V^2}{c^2}}} e_{(2)}
	\end{split}
	\label{Lorentz}
	\end{equation}
If a stationary observer sends light in a radial direction, the orbiting observer observes Doppler shift
$$
\omega'=\frac \omega {\sqrt{1 - \frac {V^2} {c^2}}}
$$
We need to add Doppler shift for gravitational field if the moving observer receives a radial wave that came from
infinity. In this case the Doppler shift will take the form
$$
\omega'=\sqrt{\frac r {r-r_g}}\frac \omega {\sqrt{ 1 - \frac {V^2} {c^2}}}
$$

We see the estimation for dynamics of star S2 that orbits Sgr A
in tables \ref{table:DShift1S2} and \ref{table:DShift2S2}.
The tables are based on two different estimations for mass of Sgr A.

If we get mass Sgr $4.1_{10}6 M_\odot$ \citeBib{Ghez} then
in pericentre (distance $1.868_{10}15$ cm) S2 has speed $738767495.4$ cm/s
and Doppler shift is $\omega'/\omega=1.000628$. In this case we measure length $2.16474 \mu m$
for emitted wave with length $2.1661 \mu m$ (Br $\gamma$).
In apocentre (distance $2.769_{10}16cm$) S2 has speed $49839993.28cm/s$
and Doppler shift is $\omega'/\omega=1.0000232$. We measure length $2.166049 \mu m$
for the same wave.
Difference between two measurements of wave length is $13.098\AA$

If we get mass Sgr $3.7_{10}6 M_\odot$ \citeBib{Schodel} then
in pericentre (distance $1.805_{10}15cm$) S2 has speed $713915922.3cm/s$
and Doppler shift is $\omega'/\omega=1.000587$. In this case we measure length $2.16483 \mu m$
for emitted wave with length $2.1661 \mu m$ (Br $\gamma$).
In apocentre (distance $2.676_{10}16cm$) S2 has speed $48163414.05cm/s$
and Doppler shift is $\omega'/\omega=1.00002171$. We measure length $2.1666052 \mu m$
for the same wave.
Difference between two measurements of wave length is $12.232\AA$

\begin{table}[h]
\caption{Doppler shift on the Earth of a wave emitted from S2;
mass of Sgr A is $4.1_{10}6 M_\odot$ \citeBib{Ghez}}
\label{table:DShift1S2}
\begin{tabular}{|c|c|c|}
\hline
&pericentre&apocentre\\
\hline
distance cm&$1.868_{10}15$&$2.769_{10}16$\\
\hline
speed cm/s&$738767495.4$&$49839993.28$\\
\hline
$\omega'/\omega$&$1.000628$&$1.0000232$\\
\hline
emitted wave $(Br\ \gamma)\ \mu m$& $2.1661$&$2.1661$\\
\hline
observed wave $\mu m$&$2.16474$ &$2.166049$\\
\hline
\end{tabular}

Difference between two measurements of wavelength is $13.098\AA$
\end{table}

\begin{table}[h]
\caption{Doppler shift on the Earth of wave emitted from S2;
mass of Sgr A is $3.7_{10}6 M_\odot$ \citeBib{Schodel}}
\label{table:DShift2S2}
\begin{tabular}{|c|c|c|}
\hline
&pericentre&apocentre\\
\hline
distance cm&$1.805_{10}15$&$2.676_{10}16$\\
\hline
speed cm/s&$713915922.3$&$48163414.05$\\
\hline
$\omega'/\omega$&$1.000587$&$1.00002171$\\
\hline
emitted wave $(Br\ \gamma)\ \mu m$& $2.1661$&$2.1661$\\
\hline
observed wave $\mu m$&$2.16483$ &$2.1666052$\\
\hline
\end{tabular}

Difference between two measurements of wavelength is $12.232\AA$
\end{table}

Difference between two measurements of wavelength in pericentre is $0.9\AA$.
Analyzing this data we can conclude that the use of Doppler shift can help improve estimation
of the mass of Sgr A.

			\section{Lorentz Transformation in Radial Direction}

We see that the Lorentz transformation in orbial direction has familiar form.
It is very interesting to see what form this transformation has for radial direction.
We start from procedure of measurement speed and use coordinate speed $v$
	\begin{equation}
dr=vdt
	\label{eq: RadialSpeed}
	\end{equation}
The time of travel of light in both directions is the same.
Trajectory of light is determined by equation $ds^2=0$.
\[\frac {r-r_g} r c^2 dt^2 - \frac r {r-r_g} dr^2 = 0\]
When light returns back to observer $A$ the change of $t$ is
\[dt=2 \frac r {r-r_g} c^{-1} dr\]
The proper time of observer $A$ is
	\begin{align*}
ds^2&= \frac {r-r_g} r c^2 4 \frac {r^2} {{r-r_g}^2}c^{-2}dr^2=\\
&=  4 \frac r {r-r_g}dr^2
	\end{align*}
Therefore spatial distance is
\[L=\sqrt{\frac r {r-r_g}} dr\]
When object moving with speed \eqref{eq: RadialSpeed} gets to $B$
change of $t$ is $\frac {dr} v$.
The proper time of observer $A$ at this point is 
\[ds^2 = \frac {r - r_g} r c^2 dr^2 v^{-2}\]
\[T = \sqrt{\frac {r - r_g} r} v^{-1} dr\]
Therefore the observer $A$ measures speed
	\begin{align*}
V&=\frac L T =
=\frac{\sqrt{\frac r {r-r_g}} dr}{\sqrt{\frac {r - r_g} r} v^{-1} dr}=\\
&=\frac r {r-r_g} v
	\end{align*}

Now we are ready to find out Lorentz transformation.
The basis vectors for stationary observer are
\[e_{(0)}=\left(\frac 1 c \sqrt{\frac r {r-r_g}}, 0, 0, 0\right)\]
\[e_{(1)}=\left(0,  \sqrt{\frac {r-r_g} r},0, 0\right)\]

Unit vector of speed should be proportional to vector
	\begin{equation}
(1, v, 0, 0)
	\label{eq: NewRadialVectorE1}
	\end{equation}
The length of this vector is
	\begin{equation}
	\begin{split}
L^2&=\frac {r-r_g} r c^2 - \frac r {r-r_g} v^2=\\
&=\frac {r-r_g} r c^2\left(1 - \frac {V^2}{c^2}\right)
	\label{eq: LengthNewRadialVectorE1}
	\end{split}
	\end{equation}
Therefore time ort of moving observer is
	\begin{align*}
e'_{(0)}&=\left(\frac 1 L,  \frac v L,0, 0\right)=\\
&=\left(\sqrt{\frac r {r-r_g}} c^{-1}\frac 1 {\sqrt{\left(1 - \frac {V^2}{c^2}\right)}}, 
v \sqrt{\frac r {r-r_g}} c^{-1}\frac 1 {\sqrt{\left(1 - \frac {V^2}{c^2}\right)}}, 0,0\right)\\
&=\left(\sqrt{\frac r {r-r_g}} c^{-1}\frac 1 {\sqrt{\left(1 - \frac {V^2}{c^2}\right)}}, 
\frac V c \sqrt{\frac {r-r_g} r}\frac 1 {\sqrt{\left(1 - \frac {V^2}{c^2}\right)}}, 0,0\right)
	\end{align*}
Spatial ort $e'_{(1)}=(A,B,0,0)$ is orthogonal $e'_{(0)}$ and has length $-1$. Therefore
	\begin{equation}
\frac {r-r_g} r c^2\frac 1 L A - \frac r {r-r_g} \frac v L B = 0
	\label{NewRadialVectorE0_1}
	\end{equation}
	\begin{equation}
\frac {r-r_g} r c^2 A^2 - \frac r {r-r_g} B^2 = -1
	\label{NewRadialVectorE0_2}
	\end{equation}
We can express $A$ from \eqref{NewRadialVectorE0_1}
\[A = c^{-2}\frac {r^2}{(r-r_g)^2} v B = c^{-2}\frac r{r-r_g} V B\]
and substitute into \eqref{NewRadialVectorE0_2}
\[\frac {r-r_g} r c^2 c^{-4}\frac {r^2}{(r-r_g)^2} V^2 B^2 - \frac r {r-r_g} B^2 = -1\]
\[\frac r{r-r_g} B^2\left(1- \frac{V^2}{c^2} \right) = 1\]
\[ B^2 = \frac{r-r_g} r\frac 1{1- \frac{V^2}{c^2} }\]
\[ B = \sqrt{\frac{r-r_g} r}\frac 1{\sqrt{1- \frac{V^2}{c^2}} }\]
	\begin{align*}
A &= c^{-2}\frac r{r-r_g} V \sqrt{\frac{r-r_g} r}\frac 1{\sqrt{1- \frac{V^2}{c^2}} }\\
&= c^{-2} V \sqrt{\frac r{r-r_g}}\frac 1{\sqrt{1- \frac{V^2}{c^2}} }
	\end{align*}
Finally spatial ort in direction of movement is
\[e'_{(1)}=\left(c^{-2} V \sqrt{\frac r{r-r_g}}\frac 1{\sqrt{1- \frac{V^2}{c^2}} },
\sqrt{\frac{r-r_g} r}\frac 1{\sqrt{1- \frac{V^2}{c^2}} },0,0\right)\]

Therefore we get transformationin in familiar form
	\begin{equation}
	\begin{split}
{e'}_{(0)}=\frac 1 {\sqrt{1 - \frac {V^2}{c^2}}} e_{(0)} +
\frac V  c \frac 1 {\sqrt{1 - \frac {V^2}{c^2}}} e_{(1)}
\\
{e'}_{(1)}=\frac V c \frac 1 {\sqrt{1 - \frac {V^2}{c^2}}} e_{(0)} +
\frac 1 {\sqrt{1 - \frac {V^2}{c^2}}} e_{(1)}
	\end{split}
	\label{RadialLorentz}
	\end{equation}

			\section{Doppler Shift in Friedman Space}

We consider another example in Friedman space. Metric of the space is
$$ds^2 = a^2 (dt^2
- d\chi^2 - \sin^2\chi (d\theta^2 - \sin^2 \theta d\phi^2))
$$
for closed model and
$$ds^2 = a^2 (dt^2
- d\chi^2 - \sinh^2\chi (d\theta^2 - \sin^2 \theta d\phi^2))
$$
for open one. Connection in this space is ($\alpha$, $\beta$ get values 1, 2, 3)
\[\Gamma^0_{00}=\frac {\dot{a}} a\]
\[\Gamma^0_{\alpha\alpha}=-\frac {\dot{a}} {a^2}g_{\alpha\alpha}\]
\[\Gamma^\alpha_{0\beta}=\frac {\dot{a}} a\delta^\alpha_\beta\]

Because space is homogenius we do not care about direction of light.
In this case
\[dk^0=-\Gamma^0_{ij}k^i k^j\]
Because $k$ is isotropic vector tangent to its trajectory we have
\[dx^\alpha=\frac {k^\alpha} {k^0}dt\]
Because $k^0=\frac \omega a$, than
\[d\frac \omega a=-\frac {da} {a^2}\omega+\frac {da} {\omega a}g_{\alpha\alpha}k^\alpha k^\alpha=
-2\frac {da} {a^2}\omega\]
\[ad\omega+\omega da=0\]
\[a\omega=const\]
Therefore when $a$ grows $\omega$ becomes smaller and length of waves grows as well. 

$a$ grows during light travel through spacetime
and this leads to red shift. We observe red shift because geometry changes,
but not because galaxies runs away one from other.

Now we want to see how red shift changes with time
if initial and final points do not move.
For simplicity I will change only $\chi$. Initial value is $\chi_1$
and final value is $\chi_2$. Because $dt=d\chi$ on light trajectory we have
\[\chi = \chi_1 + t - t_1\ \ \ \ t_2 = \chi_2 - \chi_1 + t_1\]
Therefore $a(t_1) \omega_1 = a(t_2) \omega_2$.
Doppler shift is
\[K(t_1) =\frac {\omega_2} {\omega_1} = \frac {a(t_1)} {a(t_2)}\]
If initial time changes t'1 = t1 + dt then K(t1 + dt) = a(t1 + dt) / a(t2 + dt) 
Time derivative of $K$ is
\[\dot{K} = \frac {\dot{a}_1 a_2 - a_1 \dot{a}_2} {a_2^2}\]

For closed space $a = \cosh t$. Then $\dot{a} = \sinh t$. 
\[\dot{K} = \frac{\sinh t_1 \cosh t_2 - \sinh t_2 \cosh t_1} {\cosh^2 t_2}
= \frac{\sinh(t_1 - t_2)} {\cosh^2 t_2}\] 
$K$ decreases when $t_1$ increases.

			\section{Lorentz Transformation in Friedman Space}

To learn Lorentz transformation in Friedman space I want to use metric in form
\[ds^2=c^2 dt^2 -a^2(d\chi^2 + b^2(d\theta^2 - \sin^2 \theta d\phi^2))\]
Now I have 2 observers. One does not move and has speed $(1,0,0,0)$,
and another moves along $\chi$ and his speed is $C=(1,v,0,0)$ and we assme $V=a v$.

Metric is diagonal and coordinates $\theta, \phi$ do not change.
We have transformation in plane $t, \chi$.

Unit speed of first observer is
\[e_0 = (\frac 1 c , 0, 0, 0)\]
and vector orthonormal this one is
\[e_1 = (0, \frac 1 a , 0, 0)\]

The length of vector $C$ is
\[L=\sqrt{c^2 - a^2 v^2} = c \sqrt{1 - \frac{V^2} {c^2}}\]
Therefore unit vector of speed of second observer is
\[e'_0=(\frac 1 L, \frac v L, 0, 0)\]
We look for vector
\[e'_1=(A, B, 0, 0)\]
that is orthogonal to vector $e'_0$. For this we have
	\begin{equation}
c^2 A^2 - a^2 B^2 = -1
	\label{eq: length of vector, e1, 1}
	\end{equation}
	\begin{equation}
c^2 A \frac 1 L - a^2 B \frac v L = 0
	\label{eq: length of vector, e1, 2}
	\end{equation}
We get from the equation \eqref{eq: length of vector, e1, 2}
	\begin{equation}
A = \frac {a^2} {c^2} v B
	\label{eq: length of vector, e1, 3}
	\end{equation}
We substitute \eqref{eq: length of vector, e1, 3}
into \eqref{eq: length of vector, e1, 1} and get
\[B^2 (\frac {a^4} {c^2} v^2 - a^2) = -1\] 
\[B = \frac 1 {a \sqrt{1 - \frac {V^2} {c^2}}}\]
Therefore basis of second observer is
\[e'_0=(\frac 1 {c \sqrt{1 - \frac{V^2} {c^2}}}, \frac V {c a\sqrt{1 - \frac{V^2} {c^2}}}, 0, 0)\]
\[e'_1=( \frac V {c^2 \sqrt{1 - \frac {V^2} {c^2}}},
\frac 1 {a \sqrt{1 - \frac {V^2} {c^2}}}, 0, 0)\]
Now we can express $e'$ through $e$
\[e'_0= \frac 1 {\sqrt{1 - \frac {V^2} {c^2}}} e_0
+ \frac V c \frac 1 {\sqrt{1 - \frac{V^2} {c^2}}} e_1\]
\[e'_1= \frac V c \frac 1 {\sqrt{1 - \frac{V^2} {c^2}}} e_0
+ \frac 1 {\sqrt{1 - \frac{V^2} {c^2}}} e_1\] 

\OpenBiblio


\BiblioItem{texSpaceTime}{Einstein: Geometry and Experience}
{
Einstein, Geometry and Experience, (1921)
}%

\BiblioItem{texGenRelativity}{Ghez}
{
Ghez et al.,
The First Measurement of Spectral Lines in a Short-Period Star Bound to the Galaxy's Central Black Hole: A Paradox of Youth,
\href{http://www.journals.uchicago.edu/ApJ/journal/issues/ApJL/v586n2/16990/brief/16990.abstract.html}{ApJL, 586, L127} (2003),
eprint \href{http://arxiv.org/abs/astro-ph/0302299}{arXiv:astro-ph/0302299} (2003)
}%

\BiblioItem{texGenRelativity}{Schodel}
{
R. Sch\"odel et al.,
A star in a 15.2-year orbit around the supermassive black hole at the centre of the Milky Way,
\href{http://www.nature.com/cgi-taf/DynaPage.taf?file=/nature/journal/v419/n6908/abs/nature01121_fs.html}{Nature 419, 694} (2002)
}%

\BiblioItem{texAffine,texGeomObject}{Mielke}
{
Eckehard W. Mielke, Affine generalization of the Komar complex of general relativity,
\href{http://prola.aps.org/searchabstract/PRD/v63/i4/e044018}{Phys. Rev. D 63, 044018} (2001)
}%

\BiblioItem{texAffine}{Obukhov}
{
Yu. N. Obukhov and J. G. Pereira, Metric-affine approach to teleparallel gravity,
\href{http://scitation.aip.org/getabs/servlet/GetabsServlet?prog=normal&id=PRVDAQ000067000004044016000001&idtype=cvips&gifs=Yes}
{Phys. Rev. D 67, 044016} (2003),
eprint \href{http://arxiv.org/abs/gr-qc/0212080}{arXiv:gr-qc/0212080} (2002)
}%

\BiblioItem{texAffine}{Sardanashvily}
{
Giovanni Giachetta, Gennadi Sardanashvily, Dirac Equation in Gauge and Affine-Metric Gravitation Theories,
eprint \href{http://arxiv.org/abs/gr-qc/9511035}{arXiv:gr-qc/9511035} (1995)
}%

\BiblioItem{texAffine}{Gauge}
{
Frank Gronwald and Friedrich W. Hehl, On the Gauge Aspects of Gravity, eprint
\href{http://arxiv.org/abs/gr-qc/9602013}{arXiv:gr-qc/9602013} (1996)
}%

\BiblioItem{texAffine}{Neeman}
{
Yuval Neeman, Friedrich W. Hehl, Test Matter in a Spacetime with Nonmetricity, eprint
\href{http://arxiv.org/abs/gr-qc/9604047}{arXiv:gr-qc/9604047} (1996)
}%

\BiblioItem{texTidal,texAffine,texGeomObject}{torsion}
{
F. W. Hehl, P. von der Heyde, G. D. Kerlick, and J. M. Nester,
General relativity with spin and torsion: Foundations and prospects,
\href{http://prola.aps.org/abstract/RMP/v48/i3/p393_1}{Rev. Mod. Phys. 48, 393} (1976)
}%

\BiblioItem{texTidal,texNewton}{Megged}
{
O. Megged, Post-Riemannian Merger of Yang-Mills Interactions with Gravity,
eprint \href{http://arxiv.org/abs/hep-th/0008135}{arXiv:hep-th/0008135} (2001)
}%


\BiblioItem{texNewton}{gr-qc-9604027}
{
Yu.N. Obukhov, E.J. Vlachynsky, W. Esser, R. Tresguerres and F.W. Hehl,
An exact solution of the metric-affine gauge theory with dilation, shear, and spin charges,
eprint \href{http://arxiv.org/abs/gr-qc/9604027}{arXiv:gr-qc/9604027} (1996)
}%

\BiblioItem{texLagrange}{Weinberg}
{
Steven Weinberg. The Quantum Theory of Fields. Cambridge university press.
}%

\BiblioItem{texLagrange}{Reinhardt}
{
Greiner Reinhardt. Field Quantization. Springer.
}%

\BiblioItem{texLagrange}{Landau}
{
L. D. Landau, E. M. Lifshich, The classical theory of fields.
Oxford, New York, Pergamon Press
}%

\BiblioItem{texTidal}{Wheeler}
{
Ignazio Ciufolini, John Wheeler. Gravitation and Inertia.
Princeton university press.
}%

\BiblioItem{texTidal}{Anderson02}
{
J. D. Anderson, P. A. Laing, E. L. Lau, A. S. Liu, M. M. Nieto, and S. G. Turyshev,
Study of the anomalous acceleration of Pioneer 10 and 11,
\href{http://prola.aps.org/searchabstract/PRD/v65/i8/e082004}{Phys. Rev. D 65, 082004, 50 pp.}, (2002),
eprint \href{http://arxiv.org/abs/gr-qc/0104064}{arXiv:gr-qc/0104064} (2001)
}%

\BiblioItem{texTidal}{Anderson98}
{
J. D. Anderson, P. A. Laing, E. L. Lau, A. S. Liu, M. M. Nieto, and S. G. Turyshev,
Indication, from Pioneer 10/11, Galileo, and Ulysses Data, of an Apparent Anomalous, Weak, Long-Range Acceleration,
\href{http://prola.aps.org/abstract/PRL/v81/i14/p2858_1}{Phys. Rev. Lett. 81, 2858}, (1998),
eprint \href{http://arxiv.org/abs/gr-qc/9808081}{arXiv:gr-qc/9808081} (1998)
}%


\BiblioItem{texReferenceFrame,texFiberedAlgebra}{Serge Lang}
{
Serge Lang,
Algebra, Springer, 2002
}%

\BiblioItem{texFiberedAlgebra,texTstarMorphism}{Burris Sankappanavar}
{
S. Burris, H.P. Sankappanavar,
A Course in Universal Algebra, Springer-Verlag (March, 1982),
\\eprint
\href{http://www.math.uwaterloo.ca/~snburris/htdocs/ualg.html}
{http://www.math.uwaterloo.ca/~snburris/htdocs/ualg.html}
\\(The Millennium Edition)
}%

\BiblioItem{texGeomObject}{Shilov}
{
G. E. Shilov,
Calculus, Multivariable Functions,
Moscow, Nauka, 1972
}%

\BiblioItem{texAffine,texRepresentation,texBasis,texDrcBasis,texLinearMap,texVectorSpace}{Rashevsky}
{
P. K. Rashevsky, Riemann Geometry and Tensor Calculus,\\
Moscow, Nauka, 1967
}%

\BiblioItem{texDrcBasis,texBasis}{Korn}
{
Granino A. Korn, Theresa M. Korn,
Mathematical Handbook for Scientists and Engineer,
McGraw-Hill Book Company, New York, San Francisco,
Toronto, London, Sydney, 1968
}%


\BiblioItem{texGenRelativity}{Tartaglia}
{
Angelo Tartaglia and Matteo Luca Ruggiero,
Angular Momentum Effects in Michelson\Hyph Morley Type Experiments,
Gen.Rel.Grav. 34, 1371-1382 (2002),\\
eprint \href{http://arxiv.org/abs/gr-qc/0110015}{arXiv:gr-qc/0110015} (2001)
}%

\BiblioItem{texGenRelativity}{Tomozawa}
{
Yukio Tomozawa, Speed of Light in Gravitational Fields, eprint
\href{http://arxiv.org/abs/astro-ph/0303047}{arXiv:astro-ph/0303047} (2004)
}%

\BiblioItem{texGenRelativity}{Magueijo}
{
Joao Magueijo,
Covariant and locally Lorentz-invariant varying speed of light theories,
\href{http://prola.aps.org/abstract/PRD/v62/i10/e103521}{Phys. Rev. D 62, 103521} (2000),
eprint \href{http://arxiv.org/abs/gr-qc/0007036}{arXiv:gr-qc/0007036} (2000)
}%

\BiblioItem{texGenRelativity}{Bassett}
{
Bruce A. Bassett, Stefano Liberati, Carmen Molina-Paris, and Matt Visser,
Geometrodynamics of variable-speed-of-light cosmologies,
\href{http://prola.aps.org/abstract/PRD/v62/i10/e103518}{Phys. Rev. D 62}, 103518 (2000),
eprint \href{http://arxiv.org/abs/astro-ph/0001441}{arXiv:astro-ph/0001441} (2000)
}%

\BiblioItem{texGenRelativity}{Straumann}
{
Lochlainn O'Raifeartaigh and Norbert Straumann,
Gauge theory: Historical origins and some modern developments,
\href{http://prola.aps.org/abstract/RMP/v72/i1/p1_1}{Rev. Mod. Phys. 72, 1} (2000)
}%

\BiblioItem{texGenRelativity}{Lammerzahl}
{
Claus L\"ammerzahl, Mark P. Haugan,
On the interpretation of Michelson\Hyph Morley experiments,
{Phys. Lett. A282 223-229} (2001),\\
eprint \href{http://arxiv.org/abs/gr-qc/0103052}{arXiv:gr-qc/0103052} (2001)
}%

\BiblioItem{texGenRelativity}{Muller}
{
Holger Muller et al.,
Modern Michelson-Morley Experiment using Cryogenic Optical Resonators,
\href{http://prola.aps.org/searchabstract/PRL/v91/i2/e020401}{Phys. Rev. Lett. 91, 020401} (2003),
eprint \href{http://arxiv.org/abs/physics/0305117}{arXiv:physics/0305117} (2000)
}%

\BiblioItem{texGenRelativity,texTidal}{Ranada}
{
Antonio F. Ranada,
Pioneer acceleration and variation of light speed: experimental situation,
eprint \href{http://arxiv.org/abs/gr-qc/0402120}{arXiv:gr-qc/0402120} (2004)
}%

\BiblioItem{texBiring,texVectorSpace}{math.QA-0208146}
{
I. Gelfand, S. Gelfand, V. Retakh, R. Wilson,
Quasideterminants,\\
eprint \href{http://arxiv.org/abs/math.QA/0208146}{arXiv:math.QA/0208146} (2002)
}%

\BiblioItem{texBiring,texVectorSpace}{q-alg-9705026}
{
I.Gelfand, V.Retakh,
Quasideterminants, I,\\
eprint \href{http://arxiv.org/abs/q-alg/9705026}{arXiv:q-alg/9705026} (1997)
}%

\BiblioItem{texVectorSpace}{Gelfand Retakh 1991}
{
I. Gelfand and V. Retakh, Determinants of Matrices over Noncommutative Rings, Funct.
Anal. Appl. 25 (1991), no. 2, 91-102
}%

\BiblioItem{texVectorSpace}{Gelfand Retakh 1992}
{
I. Gelfand and V. Retakh, A Theory of Noncommutative Determinants and Characteristic
Functions of Graphs, Funct. Anal. Appl. 26 (1992), no. 4, 1-20
}%

\BiblioItem{texVectorSpace}{hep-th-9407124}
{
I. M. Gelfand, D. Krob, A. Lascoux, B. Leclerc, V.S. Retakh and J.-Y. Thibon,
Noncommutative symmetric functions,\\
eprint \href{http://arxiv.org/abs/hep-th/9407124}{arXiv:hep-th/9407124} (1994)
}%

\BiblioItem{texVectorSpace}{Carl Faith 1}
{
Carl Faith, Algebra: Rings, Modules and Categories I,
Springer - Verlag, Berlin - Heidelberg - New York, 1973
}%



\BiblioItem{texDrcReferenceFrame,texRefernceFrame,texGeomObject,texLie,texLieRepresentation}{0412.391}
{
Aleks Kleyn,
Basis Manifold,
eprint \href{http://arxiv.org/abs/math.DG/0412391}{arXiv:math.DG/0412391} (2004)
}%

\BiblioItem{texAffine}{0405.027}
{
Aleks Kleyn,
Reference Frame in General Relativity,
eprint \href{http://arxiv.org/abs/gr-qc/0405027}{arXiv:gr-qc/0405027} (2004)
}%

\BiblioItem{texTidal}{0405.028}
{
Aleks Kleyn, Metric-Affine Manifold,
eprint \href{http://arxiv.org/abs/gr-qc/0405028}{arXiv:gr-qc/0405028} (2004)
}%

\BiblioItem{texFiberedAlgebra,texBundleRelation,texTstarMorphism}{0701.238}
{
Aleks Kleyn,
Lectures on Linear Algebra over Skew Field,\\
eprint \href{http://arxiv.org/abs/math.GM/0701238}{arXiv:math.GM/0701238} (2007)
}%

\BiblioItem{texBundleRelation,texPrefaceRelation}{0702.561}
{
Aleks Kleyn,
Algebra Bundle,\\
eprint \href{http://arxiv.org/abs/math.DG/0702561}{arXiv:math.DG/0702561} (2007)
}%


\BiblioItem{texPolymodule}{math.RA-0501237v1}
{
Aleks Kleyn,
Module Over Skew-Field, version 1,\\
eprint \href{http://arxiv.org/abs/math/0501237v1}{arXiv:math.RA/0501237v1} (2005)
}%

\ifx\texBiring\Defined
\else
\BiblioItem{texVectorSpace,texFiberedAlgebra}{0612.111}
{
Aleks Kleyn,
Biring of Matrices,\\
eprint \href{http://arxiv.org/abs/math.OA/0612111}{arXiv:math.OA/0612111} (2006)
}%
\fi

\ifx\texBundleRelation\Defined
\else
\BiblioItem{texFiberedMorphism}{0707.2246}
{
Aleks Kleyn,
Fibered Correspondence,\\
eprint \href{http://arxiv.org/abs/0707.2246}{arXiv:0707.2246} (2007)
}%
\fi

\ifx\PrintBook\Defined
\BiblioItem{texPrefaceRelation}{0707.2246}
{
Aleks Kleyn,
Fibered Correspondence,\\
eprint \href{http://arxiv.org/abs/0707.2246}{arXiv:0707.2246} (2007)
}%
\fi

\BiblioItem{texHomotopy}{q-alg-9705009}
{
John C. Baez,
An Introduction to n-Categories,\\
eprint \href{http://arxiv.org/abs/q-alg/9705009}{arXiv:q-alg/9705009} (1997)
}%

\BiblioItem{texSpaceTime}{Einstein: Isaak Newton}
{
Albert Einstein,
Isaak Newton,
Manchester Guardian, 16, 234 - 235 (1927)
}%

\BiblioItem{texPrefaceRelation}{Tolstoi about Anna Karenina}
{
Tolstoi about Anna Karenina,
in book A Karenina Companion, by C. J. G. Turner,
published by Wilfrid Laurier University Press (August 1993)
}%

\BiblioItem{texBundleRelation,texPrefaceRelation,texTstarMorphism,texBundle}
{Cohn: Universal Algebra}
{
Paul M. Cohn,
Universal Algebra,
Springer, 1981
}%

\BiblioItem{texBundle}
{Maunder: Algebraic Topology}
{
C. R. F. Maunder,
Algebraic Topology,
Dover Publications, Inc, Mineola, New York, 1996
}%

\BiblioItem{texFiberedAlgebra}{Pommaret: Partial Differential Equations}
{
J.-F. Pommaret,
Partial Differential Equations and Group Theory,
Springer, 1994
}%

\BiblioItem{texBundleRelation}{Bourbaki: Set Theory}
{
N. Bourbaki,
Theory of sets,
Springer, 2004
}%

\BiblioItem{texBundle,texCartesian,texFiberedAlgebra,texBundleRelation,texFiberedMorphism}
{Bourbaki: General Topology 1}
{
N. Bourbaki,
General Topology, Chapters 1 - 4,
Springer, 1989
}

\BiblioItem{texCalculus}{Bourbaki: General Topology: Chapter 5 - 10}
{
N. Bourbaki,
General Topology, Chapters 5 - 10,
Springer, 1989
}

\BiblioItem{texCalculus}{Bourbaki: Topological Vector Space}
{
N. Bourbaki,
Topological Vector Spaces, Chapters 1 - 5,
Springer, 1987
}

\BiblioItem{texCalculus}{Pontryagin: Topological Group}
{
L. S. Pontryagin,
Selected Works, Volume Two, Topological Groups,
Gordon and Breach Science Publishers, 1986
}

\BiblioItem{texFiberedMorphism}{Postnikov: Differential Geometry}
{
Postnikov M. M.,
Geometry IV: Differential geometry,
Moscow, Nauka, 1983
}

\BiblioItem{texFiberedAlgebra,texFiberedMorphism}{Hatcher: Algebraic Topology}
{
Allen Hatcher,
Algebraic Topology,
Cambridge University Press, 2002
}

\BiblioItem{texFiberedMorphism}{geometry of differential equations}
{
Vinogradov, A. M., Krasil'shchik, I. S., and Lychagin, V. V.,
Introduction to geometry of nonlinear differential equations,
Nauka, Moscow, 1986
}

\BiblioItem{texFiberedMorphism}{cohomological analysis}
{
A. M. Vinogradov,
Cohomological Analysis of Partial Differential Equations
and Secondary Calculus,
American Mathematical Society, 2001
}

\CloseBiblio

\OpenIndex
\SetIndexSpace%
\Index{texLinearMap}
   {$1$-\drc form}%
   {1-drc form, vector spaces}%
\SetIndexSpace%
\Index{texPolymodule}
   {$(2)$\hyph vector space}%
   {(2)-vector space}%
\Index{texBundleRelation}
   {$2$\Hyph ary fibered relation}%
   {2 ary fibered relation}%
\SetIndexSpace%
\Index{texCalculus}
   {$A$\Hyph valued function}%
   {A valued function}%
\Index{texBiring}
   {$(^{\gi a}_{\gi b})$\hyph \CR quasideterminant}%
   {a b cr-quasideterminant}%
\Index{texBiring}
   {$(^{\gi a}_{\gi b})$\hyph \RC quasideterminant}%
   {a b RC-quasideterminant}%
\Index{texCalculus}
   {absolute value on skew field}%
   {absolute value on skew field}%
\Index{texBasis}
   {active representation}%
   {active representation}%
\Index{texDrcBasis}
   {active \sT representation}%
   {active representation, vector space}%
\Index{texBasis}
   {active transformation on basis manifold}%
   {active transformation}%
\Index{texBasis}
   {affine basis}%
   {Affine Basis}%
\Index{texBasis}
   {affine transformation group}%
   {AffineTransformationGroup}%
\Index{texTypeBasis}
   {affine transformation group}%
   {AffineTransformationGroup}%
\Index{texBasis}
   {affine transformation on basis manifold}%
   {affine transformation}%
\Index{texBiring}
   {alternative representation of matrix}%
   {Alternative representation}%
\Index{texRefernceFrame}
   {anholonomic coordinate}%
   {anholonomic coordinate}%
\Index{texRefernceFrame}
   {anholonomic coordinates of connection}%
   {anholonomic coordinates of connection}%
\Index{texRefernceFrame}
   {anholonomic coordinates of vector}%
   {vector anholonomic coordinates}%
\Index{texRefernceFrame}
   {anholonomic coordinates on manifold}%
   {anholonomic coordinates on manifold}%
\Index{texRefernceFrame}
   {anholonomity object}%
   {anholonomity object}%
\Index{texFiberedGroup}
   {antihomomorphism of fibered groups}%
   {antihomomorphism of fibered groups}%
\Index{texBundleRelation}
   {antisymmetric $2$\Hyph ary fibered relation}%
   {antisymmetric 2 ary fibered relation}%
\Index{texFiberedAlgebra}
   {arity of operation}%
   {arity of operation}%
\Index{texTstarRepresentation}
   {associative law for covariant \Ts representation}%
   {associative law for Tstar covariant representation}%
\Index{texDrcMorphism}
   {associative law for \drc linear maps of vector spaces}%
   {associative law for drc linear maps of vector spaces}%
\Index{texVectorSpace}
   {associative law for \Ts vector space}%
   {associative law, Tstar vector space}%
\Index{texLinearMap}
   {associative law for twin representations}%
   {associative law for twin representations}%
\Index{texBundleRelation}
   {associative law of composition of fibered correspondences}%
   {associative law, composition of fibered correspondences}%
\Index{texAffine}
   {auto parallel line}%
   {auto parallel line}%
\SetIndexSpace%
\Index{texBundleRelation}
   {base of fibered correspondence}%
   {base of fibered correspondence}%
\Index{texBundle}
   {base of map}%
   {base of map}%
\Index{texTypeBasis}
   {basis}%
   {}%
\SubIndex{texTypeBasis}
   {affine}%
   {Affine Basis}%
\SubIndex{texTypeBasis}
   {central affine}%
   {Central Affine Basis}%
\SubIndex{texTypeBasis}
   {orthonornal}%
   {Orthonornal Basis}%
\Index{texDrcBasis}
   {basis manifold}%
   {}%
\SubIndex{texTypeBasis}
   {of affine space}%
   {Basis Manifold, Affine Space}%
\SubIndex{texTypeBasis}
   {of central affine space}%
   {Basis Manifold, Central Affine Space}%
\SubIndex{texDrcBasis}
   {of \drc vector space}%
   {basis manifold of vector space}%
\SubIndex{texTypeBasis}
   {of Euclid space}%
   {Basis Manifold, Euclid Space}%
\Index{texBasis}
   {basis manifold of affine space}%
   {Basis Manifold, Affine Space}%
\Index{texBasis}
   {basis manifold of central affine space}%
   {Basis Manifold, Central Affine Space}%
\Index{texBasis}
   {basis manifold of Euclid space}%
   {Basis Manifold, Euclid Space}%
\Index{texBasis}
   {basis manifold of vector space}%
   {basis manifold of vector space}%
\Index{texBasis}
   {basis of vector space}%
   {Basis}%
\Index{texLieRepresentation}
   {basis vector}%
   {}%
\SubIndex{texLieRepresentation}
   {of \sT representation}%
   {basis vector of starT representation}%
\SubIndex{texLieRepresentation}
   {of \Ts representation}%
   {basis vector of Tstar representation}%
\Index{texBiring}
   {biring}%
   {biring}%
\Index{texFiberedMorphism}
   {bundle of level $2$}%
   {bundle of level 2}%
\Index{texFiberedMorphism}
   {bundle of level $n$}%
   {bundle of level n}%
\SetIndexSpace%
\Index{texVectorSpace}
   {\subs rows \dcr vector space}%
   {subs rows dcr vector space}%
\Index{texBiring}
   {\subs row of matrix}%
   {c row}%
\Index{texBiring}
   {$c$\hyph row of matrix}%
   {c-row}%
\Index{texAffine}
   {Cartan connection}%
   {Cartan connection}%
\Index{texAffine}
   {Cartan curvature}%
   {Cartan curvature}%
\Index{texAffine}
   {Cartan derivative}%
   {Cartan derivative}%
\Index{texAffine}
   {Cartan symbol}%
   {Cartan symbol}%
\Index{texAffine}
   {Cartan transport}%
   {Cartan transport}%
\Index{texBundle}
   {Cartesian power $\mathcal{A}$ of bundle $\mathcal{B}$}%
   {Cartesian power of bundle}%
\Index{texBundle}
   {Cartesian power $A$ of set $B$}%
   {Cartesian power of set}%
\Index{texCartesian}
   {Cartesian power of bundle}%
   {Cartesian power of bundle}%
\Index{texCartesian}
   {direct product of bundles}%
   {Cartesian product of bundles}%
\Index{texCartesian}
   {direct product of total spaces}%
   {Cartesian product of total spaces}%
\Index{texHomotopy}
   {category of \drc vector spaces}%
   {category of drc vector spaces}%
\Index{texBundleRelation}
   {category of fibered correspondences over diagonal}%
   {category of fibered correspondences over diagonal}%
\Index{texBundleRelation}
   {category of reduced fibered correspondences}%
   {category of reduced fibered correspondences}%
\Index{texBasis}
   {central affine basis}%
   {Central Affine Basis}%
\Index{texRepresentation}
   {column vector}%
   {column vector}%
\Index{texBundleRelation}
   {commutative diagram of correspondences}%
   {commutative diagram of correspondences}%
\Index{texBundle}
   {compact\hyph open topology}%
   {compact open topology}%
\Index{texDiffEq}
   {completely integrable system}%
   {completely integrable system}%
\Index{texBundleRelation}
   {composition of fibered correspondences}%
   {composition of fibered correspondences}%
\SubIndex{texVectorSpace}
   {of \drc linear equations}%
   {extended matrix, system of drc linear equations}%
\SubIndex{texVectorSpace}
   {of \rcd linear equations}%
   {extended matrix, system of rcd linear equations}%
\Index{texBundleRelation}
   {composition of reduced fibered correspondences}%
   {composition of reduced fibered correspondences}%
\Index{texBiring}
   {condition of reducibility of products}%
   {condition of reducibility of products}%
\Index{texBundleRelation}
   {continuous correspondence}%
   {continuous correspondence}%
\Index{texFiberedGroup}
   {contravariant \Ts representation of fibered group}%
   {Tstar contravariant representation of fibered group}%
\Index{texBasis}
   {coordinate isomorphism}%
   {coordinate isomorphism}%
\Index{texVectorSpace}
   {coordinate isomorphism}%
   {coordinate isomorphism}%
\Index{texVectorSpace}
   {coordinate matrix}%
   {}%
\SubIndex{texVectorSpace}
   {of set of vectors in \dcr rows vector space}%
   {coordinate matrix of set of vectors, dcr vector space}%
\SubIndex{texVectorSpace}
   {of set of vectors in \drc rows vector space}%
   {coordinate matrix of set of vectors, drc vector space}%
\SubIndex{texVectorSpace}
   {of vector in \drc basis}%
   {coordinate matrix of vector in drc basis}%
\Index{texRefernceFrame}
   {coordinate reference frame}%
   {coordinate reference frame}%
\Index{texDrcBasis}
   {coordinate representation in \drc vector space}%
   {coordinate representation, vector space}%
\Index{texBasis}
   {coordinate representation of group in vector space}%
   {coordinate representation, vector space}%
\Index{texBasis}
   {coordinate vector space}%
   {coordinate vector space}%
\Index{texBasis}
   {coordinates of geometrical object}%
   {coordinates of geometrical object, vector space}%
\Index{texDrcBasis}
   {coordinates of geometrical object}%
   {}%
\SubIndex{texDrcBasis}
   {in coordinate vector space}%
   {coordinates of geometrical object, coordinate vector space}%
\SubIndex{texDrcBasis}
   {in vector space}%
   {coordinates of geometrical object, vector space}%
\Index{texBasis}
   {coordinates of geometrical object in coordinate representation}%
   {coordinates of geometrical object, coordinate vector space}%
\Index{texBasis}
   {coordinates of representation}%
   {coordinates of representation}%
\Index{texDrcBasis}
   {coordinates of representation}%
   {coordinates of representation}%
\Index{texVectorSpace}
   {coordinates of set of vectors in \dcr vector space}%
   {coordinates of set of vectors, dcr vector space}%
\Index{texVectorSpace}
   {coordinates of set of vectors in \drc vector space}%
   {coordinates of set of vectors, drc vector space}%
\Index{texVectorSpace}
   {coordinates of vector in \drc basis}%
   {coordinates of vector in drc basis}%
\Index{texBundleRelation}
   {correspondence continuous on the set}%
   {correspondence continuous on the set}%
\Index{texBundleRelation}
   {correspondence of homomorphism}%
   {correspondence of homomorphism}%
\Index{texFiberedGroup}
   {covariant \Ts representation of fibered group}%
   {Tstar covariant representation of fibered group}%
\Index{texBiring}
   {\CR inverse element of biring}%
   {cr-inverse element}%
\Index{texVectorSpace}
   {\CR matrix group}%
   {cr-matrix group}%
\Index{texBiring}
   {\CR power}%
   {cr power}%
\Index{texBiring}
   {\CR product of matrices}%
   {cr-product of matrices}%
\Index{texVectorSpace}
   {\crd vector space}%
   {crd vector space}%
\SetIndexSpace%
\Index{texCalculus}
   {$D$\Hyph valued variable}%
   {D valued variable}%
\Index{texVectorSpace}
   {\dcr basis of \subs rows vector space}%
   {dcr basis, c rows vector space}%
\Index{texVectorSpace}
   {\dcr vector}%
   {dcr vector}%
\Index{texVectorSpace}
   {\dcr vector space}%
   {dcr vector space}%
\Index{texBiring}
   {determinant of matrix}%
   {determinant}%
\Index{texTidal}
   {deviation of trajectories}%
   {deviation of trajectories}%
\Index{texBundleRelation}
   {diagonal in bundle}%
   {diagonal in bundle}%
\Index{texBundleRelation}
   {diagram of correspondences}%
   {diagram of correspondences}%
\Index{texCalculus}
   {differentiable functions of \drc vector space to skew field $D$}%
   {differentiable functions, drc vector space to skew field}%
\Index{texCalculus}
   {differential of mapping of normed \drc vector space to valued skew field}%
   {differential, drc vector space to skew field}%
\Index{texVectorSpace}
   {dimension of \drc vector space}%
   {dimension of vector space}%
\Index{texFiberedGroup}
   {direct product of representations of fibered group}%
   {direct product of representations of fibered group}%
\Index{texRepresentation}
   {direct product of representations of group}%
   {direct product of representations of group}%
\Index{texTstarRepresentation}
   {direct product of \Ts representations of group}%
   {direct product of representations of group}%
\Index{texLieRepresentation}
   {direct sum of representations}%
   {direct sum of representations}%
\Index{texVectorSpace}
   {distributive law}%
   {}%
\SubIndex{texVectorSpace}
   {\Ts vector space}%
   {distributive law, Tstar vector space}%
\Index{texVectorSpace}
   {\drc automorphism of vector space}%
   {automorphism of vector space}%
\Index{texVectorSpace}
   {\drc basis}%
   {}%
\SubIndex{texVectorSpace}
   {for \sups rows vector space}%
   {drc basis, r rows vector space}%
\SubIndex{texVectorSpace}
   {for vector space}%
   {drc basis, vector space}%
\Index{texVectorSpace}
   {\drc coordinate vector space}%
   {drc coordinate vector space}%
\Index{texVectorSpace}
   {\drc isomorphism of vector spaces}%
   {isomorphism of vector spaces}%
\Index{texDrcMorphism}
   {\drc linear map of vector spaces}%
   {drc linear map of vector spaces}%
\Index{texVectorSpace}
   {\drc linear span in vector space}%
   {linear span, vector space}%
\Index{texVectorSpace}
   {\drc linearly dependent vectors}%
   {linearly dependent, vector space}%
\Index{texVectorSpace}
   {\drc linearly independent vectors}%
   {linearly independent, vector space}%
\Index{texVectorSpace}
   {\drc system of linear equations}%
   {system of linear equations}%
\Index{texVectorSpace}
   {\drc vector}%
   {drc vector}%
\Index{texCalculus}
   {\drc vector function}%
   {drc vector function}%
\Index{texVectorSpace}
   {\drc vector space}%
   {drc vector space}%
\Index{texVectorSpace}
   {$D\star$\hyph  vector space}%
   {Dstar vector space}%
\Index{texVectorSpace}
   {$D\star$\hyph product of vector over scalar}%
   {Dstar product of vector over scalar, vector space}%
\Index{texBiring}
   {duality principle for biring}%
   {duality principle for biring}%
\Index{texBiring}
   {duality principle for biring of matrices}%
   {duality principle for biring of matrices}%
\SetIndexSpace%
\Index{texTstarMorphism}
   {effective representation of $\mathfrak{F}$\Hyph algebra $A$}%
   {effective representation of algebra}%
\Index{texFiberedAlgebra}
   {effective representation of fibered $\mathfrak{F}$\Hyph algebra}%
   {effective representation of fibered F-algebra}%
\Index{texFiberedGroup}
   {effective \Ts representation of fibered group}%
   {effective representation of fibered group}%
\Index{texRepresentation}
   {effective representation of group}%
   {effective representation of group}%
\Index{texVectorSpace}
   {effective representation of skew field}%
   {effective representation of skew field}%
\Index{texELie}
   {enhanced Lie group}%
   {enhanced Lie group}%
\Index{texDiffEq}
   {essential parameters}%
   {essential parameters}%
\Index{texBundleRelation}
   {extension of correspondence}%
   {extension of correspondence}%
\Index{texAffine}
   {extreme line}%
   {extreme line}%
\SetIndexSpace%
\Index{texBundleRelation}
   {fibered correspondence from $\mathcal{A}$ to $\mathcal{B}$}%
   {fibered correspondence from A to B}%
\Index{texBundleRelation}
   {fibered correspondence in $\mathcal{A}$}%
   {fibered correspondence in A}%
\Index{texBundleRelation}
   {fibered correspondence of homomorphism}%
   {fibered correspondence of homomorphism}%
\Index{texBundleRelation}
   {fibered equivalence}%
   {fibered equivalence}%
\Index{texFiberedAlgebra}
   {fibered $\mathfrak{F}$\Hyph algebra}%
   {fibered F-algebra}%
\Index{texFiberedAlgebra}
   {fibered $\mathfrak{F}$\Hyph subalgebra}%
   {fibered F-subalgebra}%
\Index{texFiberedAlgebra}
   {fibered group}%
   {fibered group}%
\Index{texFiberedMorphism}
   {fibered identification morphism}%
   {fibered identification morphism}%
\Index{texFiberedMorphism}
   {fibered little group}%
   {fibered little group}%
\Index{texBundle}
   {fibered morphism from bundle $\mathcal{A}$ into $\mathcal{B}$}%
   {fibered morphism from A into B}%
\Index{texFiberedMorphism}
   {fibered natural morphism}%
   {fibered natural morphism}%
\Index{texBundleRelation}
   {fibered ordering}%
   {fibered ordering}%
\Index{texBundleRelation}
   {fibered preordering}%
   {fibered preordering}%
\Index{texFiberedAlgebra}
   {fibered ring}%
   {fibered ring}%
\Index{texFiberedMorphism}
   {fibered stability group}%
   {fibered stability group}%
\Index{texBundle}
   {fibered subset}%
   {fibered subset}%
\Index{texNewton}
   {field-strength tensor}%
   {field-strength tensor}%
\Index{texBundleRelation}
   {filter $\mathfrak{F}$ converges to $A$}%
   {filter converges}%
\Index{texNewton}
   {first Newton law}%
   {First Newton law}%
\Index{texFiberedMorphism}
   {free \Ts representation of fibered group}%
   {free representation of fibered group}%
\Index{texTstarRepresentation}
   {free \Ts representation of group}%
   {free representation of group}%
\Index{texAffine}
   {Frenet transport}%
   {Frenet transport}%
\Index{texCalculus}
   {function continuous with respect to set of arguments}%
   {function continuous with respect to set of arguments}%
\Index{texCalculus}
   {function of $\gi n$ $D$\Hyph valued variables}%
   {function of n D valued variables}%
\SetIndexSpace%
\Index{texRefernceFrame}
   {$G$\Hyph reference frame}%
   {G reference frame}%
\Index{texTypeBasis}
   {\Gbasis}%
   {G-basis}%
\Index{texBasis}
   {\Gbasis\ of vector space}%
   {G-basis}%
\Index{texBasis}
   {\Gcoords\ of basis}%
   {G-coordinates}%
\Index{texTypeBasis}
   {\Gcoords}%
   {G-coordinates}%
\Index{texTypeBasis}
   {\Gspace}%
   {GSpace}%
\Index{texDrcBasis}
   {geometrical object}%
   {}%
\SubIndex{texDrcBasis}
   {defined in vector space}%
   {geometrical object, vector space}%
\SubIndex{texDrcBasis}
   {in coordinate representation defined in vector space}%
   {geometrical object, coordinate vector space}%
\SubIndex{texDrcBasis}
   {of type $A$}%
   {geometrical object of type A, vector space}%
\Index{texBasis}
   {geometrical object in coordinate representation}%
   {geometrical object, coordinate vector space}%
\Index{texBasis}
   {geometrical object in vector space}%
   {geometrical object, vector space}%
\Index{texBasis}
   {geometrical object of type $A$ in vector space}%
   {geometrical object of type A, vector space}%
\Index{texGroupRing}
   {group algebra}%
   {group algebra}%
\Index{texBasis}
   {\Gspace}%
   {GSpace}%
\SetIndexSpace%
\Index{texBiring}
   {Hadamard inverse of matrix}%
   {Hadamard inverse of matrix}%
\Index{texRefernceFrame}
   {holonomic coordinates of connection}%
   {holonomic coordinates of connection}%
\Index{texRefernceFrame}
   {holonomic coordinates of vector}%
   {vector holonomic coordinates}%
\Index{texFiberedGroup}
   {homogeneous bundle of fibered group}%
   {homogeneous bundle of fibered group}%
\Index{texTstarRepresentation}
   {homogeneous space of group}%
   {homogeneous space of group}%
\Index{texRepresentation}
   {homogeneous space of group}%
   {homogeneous space of group}%
\Index{texFiberedAlgebra}
   {homomorphism of fibered $\mathfrak{F}$\Hyph algebras}%
   {homomorphism of fibered F-algebras}%
\Index{texFiberedGroup}
   {homomorphism of fibered groups}%
   {homomorphism of fibered groups}%
\SetIndexSpace%
\Index{texLieRepresentation}
   {infinitesimal generator}%
   {infinitesimal generator}%
\Index{texLinearLie}
   {infinitesimal generators of group Lie}%
   {infinitesimal generators of group Lie}%
\Index{texDrcBasis}
   {invariance principle}%
   {invariance principle}%
\Index{texBasis}
   {invariance principle in vector space}%
   {invariance principle, vector space}%
\Index{texBundleRelation}
   {inverse fibered correspondence}%
   {inverse fibered correspondence}%
\Index{texBundleRelation}
   {inverse reduced fibered correspondence}%
   {inverse reduced fibered correspondence}%
\Index{texFiberedAlgebra}
   {isomorphism of fibered $\mathfrak{F}$\Hyph algebras}%
   {isomorphism of fibered F-algebras}%
\SetIndexSpace%
\Index{texFiberedGroup}
   {kernel of inefficiency of representation of fibered group}%
   {kernel of inefficiency of representation of fibered group}%
\Index{texRepresentation}
   {kernel of inefficiency of representation of group}%
   {kernel of inefficiency of representation of group}%
\Index{texTstarRepresentation}
   {kernel of inefficiency of \Ts representation of group $G$}%
   {kernel of inefficiency of representation of group}%
\Index{texDiffProperty}
   {Killing equation}%
   {Killing equation}%
\Index{texDiffProperty}
   {Killing equation of second type}%
   {Killing equation second type}%
\Index{texDiffProperty}
   {Killing vector of second type}%
   {Killing vector second type}%
\Index{texBiring}
   {Kronecker symbol}%
   {Kronecker symbol}%
\SetIndexSpace%
\Index{texLie}
   {left invariant vector field}%
   {left invariant vector}%
\Index{texVectorSpace}
   {left module}%
   {left module}%
\Index{texTstarRepresentation}
   {left shift}%
   {left shift}%
\Index{texFiberedGroup}
   {left shift on fibered group}%
   {Tstar shift, fibered group}%
\Index{texRepresentation}
   {left shift on group}%
   {left shift, group}%
\Index{texLie}
   {left structural constant of Lie algebra}%
   {left structural constant of Lie algebra}%
\Index{texVectorSpace}
   {left vector space}%
   {left vector space}%
\Index{texRepresentation}
   {left-side contravariant representation of group}%
   {left-side contravariant representation}%
\Index{texRepresentation}
   {left-side covariant representation of group}%
   {left-side covariant representation}%
\Index{texTstarMorphism}
   {left-side representation of $\mathfrak{F}$\Hyph algebra $A$ in set $M$}%
   {left-side representation of algebra}%
\Index{texFiberedAlgebra}
   {left-side representation of fibered $\mathfrak{F}$\Hyph algebra}%
   {left-side representation of fibered F-algebra}%
\Index{texRepresentation}
   {left-side representation of group}%
   {left-side representation of group}%
\Index{texTstarMorphism}
   {left-side transformation}%
   {left-side transformation}%
\Index{texFiberedAlgebra}
   {left-side transformation on bundle}%
   {left-side transformation of bundle}%
\Index{texLie}
   {Lie algebra of Lie group}%
   {algebra Lie group Lie}%
\SubIndex{texLie}
   {left defined}%
   {left defined algebra Lie}%
\SubIndex{texLie}
   {right defined}%
   {right defined algebra Lie}%
\Index{texDiffProperty}
   {Lie derivative}%
   {Lie derivative}%
\SubIndex{texDiffProperty}
   {of connection}%
   {Lie derivative of connection}%
\SubIndex{texDiffProperty}
   {of metric}%
   {Lie derivative of metric}%
\Index{texLie}
   {Lie group basic operators}%
   {Lie group basic operators}%
\Index{texBundleRelation}
   {lift of correspondence}%
   {lift of correspondence}%
\Index{texBundle}
   {lift of map}%
   {lift of map}%
\Index{texBundleRelation}
   {limit of correspondence with respect to the filter}%
   {limit of correspondence with respect to the filter}%
\Index{texBundleRelation}
   {limit of filter}%
   {limit of filter}%
\Index{texBundleRelation}
   {limit set of filter}%
   {limit set of filter}%
\Index{texRepresentation}
   {linear representation of group}%
   {linear representation of group}%
\Index{texTstarRepresentation}
   {little group}%
   {little group}%
\Index{texRefernceFrame}
   {local reference frame}%
   {local reference frame}%
\Index{texRefernceFrame}
   {Lorentz transformation}%
   {Lorentz transformation}%
\SetIndexSpace%
\Index{texDrcReferenceFrame}
   {map of type $G$ on manifold}%
   {map of type G on manifold}%
\Index{texCartesian}
   {mapping space}%
   {mapping space}%
\Index{texDrcMorphism}
   {matrix of \drc linear map}%
   {matrix of drc linear map}%
\Index{texGeomObject}
   {metric-affine manifold}%
   {metric-affine manifold}%
\Index{texFiberedMorphism}
   {morphism of fibered \Ts representations from $\mathcal{F}$ into $\mathcal{G}$}%
   {morphism of fibered representations from f into g}%
\Index{texTstarMorphism}
   {morphism of \Ts representations from $f$ into $g$}%
   {morphism of representations from f into g}%
\Index{texTstarMorphism}
   {morphism of \Ts representations of $\mathfrak{F}$ algebra}%
   {morphism of representations of F algebra}%
\Index{texTstarMorphism}
   {morphism of \Ts representations of $\mathfrak{F}$\Hyph algebra in $\mathfrak{H}$\Hyph algebra}%
   {morphism of representations of F algebra in H algebra}%
\Index{texFiberedMorphism}
   {morphism of \Ts representations of fibered $\mathfrak{F}$\Hyph algebra}%
   {morphism of representations of fibered F algebra}%
\Index{}
   {movement on basis manifold}%
   {movement transformation}%
\SetIndexSpace%
\Index{texPolymodule}
   {$(n)$\hyph vector space}%
   {(n)-vector space}%
\Index{texBundleRelation}
   {$n$\Hyph ary fibered relation}%
   {fibered relation}%
\Index{texGeomObject}
   {nonmetricity}%
   {nonmetricity}%
\Index{texVectorSpace}
   {nonsingular \drc system of linear equations}%
   {nonsingular system of linear equations}%
\Index{texRepresentation}
   {nonsingular \Ts transformation}%
   {nonsingular transformation}%
\Index{texCalculus}
   {norm on \drc vector space}%
   {norm on drc vector space}%
\Index{texCalculus}
   {normed \drc vector space}%
   {normed drc vector space}%
\SetIndexSpace%
\Index{texFiberedAlgebra}
   {operation on bundle}%
   {operation on bundle}%
\Index{texBundleRelation}
   {opposite fibered preordering}%
   {opposite fibered preordering}%
\Index{texFiberedGroup}
   {orbit of representation of fibered group}%
   {orbit of representation of fibered group}%
\Index{texRepresentation}
   {orbit of representation of group}%
   {orbit of representation of group}%
\Index{texTstarRepresentation}
   {orbit of \Ts representation of group}%
   {orbit of representation of group}%
\Index{texBasis}
   {orthonornal basis}%
   {Orthonornal Basis}%
\SetIndexSpace%
\Index{texGeomObject}
   {parallelogram}%
   {parallelogram}%
\Index{texCalculus}
   {partial derivative of mapping $f$ with respect to variable $v^{\gi a}$}%
   {partial derivative of mapping with respect to variable, skew field}%
\Index{texCalculus}
   {partial derivative of mapping $\Vector f$ with respect to variable $v^{\gi a}$}%
   {partial derivative of mapping with respect to variable, drc vector space}%
\Index{texBasis}
   {passive representation}%
   {passive representation}%
\Index{texBasis}
   {passive transformation on basis manifold}%
   {passive transformation}%
\Index{texDrcBasis}
   {passive \Ts representation}%
   {passive representation}%
\Index{texRefernceFrame}
   {pfaffian derivative}%
   {pfaffian derivative}%
\Index{texNewton}
   {potential energy}%
   {potential energy}%
\Index{texDrcBasis}
   {product of geometrical object and constant}%
   {product of geometrical object and constant}%
\Index{texBasis}
   {product of geometrical object and constant in vector space}%
   {product of geometrical object and constant, vector space}%
\Index{texTstarMorphism}
   {product of morphisms of \Ts representations of $\mathfrak{F}$\Hyph algebra}%
   {product of morphisms of representations of F algebra}%
\Index{texBundle}
   {projection of bundle $\mathcal{E}$ along fiber $E$}%
   {projection of bundle along fiber}%
\SetIndexSpace%
\Index{texBasis}
   {quasi affine transformation on basis manifold}%
   {quasi affine transformation}%
\Index{texBasis}
   {quasi movement on basis manifold}%
   {quasi movement}%
\Index{texFiberedMorphism}
   {quotient bundle}%
   {quotient bundle}%
\SetIndexSpace%
\Index{texBiring}
   {\sups row of matrix}%
   {r row}%
\Index{texVectorSpace}
   {\sups rows \drc vector space}%
   {sups rows drc vector space}%
\Index{texBiring}
   {$r$\hyph row of matrix}%
   {r-row}%
\Index{texBiring}
   {\RC inverse element of biring}%
   {rc-inverse element}%
\Index{texVectorSpace}
   {\RC major minor}%
   {RC-major minor}%
\Index{texVectorSpace}
   {\RC matrix group}%
   {rc-matrix group}%
\Index{texVectorSpace}
   {\RC nonsingular matrix}%
   {RC nonsingular matrix}%
\Index{texBiring}
   {\RC power}%
   {rc power}%
\Index{texBiring}
   {\RC product of matrices}%
   {rc-product of matrices}%
\Index{texBiring}
   {\RC quasideterminant}%
   {RC-quasideterminant}%
\Index{texVectorSpace}
   {\RC rank of matrix}%
   {rc-rank of matrix}%
\Index{texVectorSpace}
   {\RC singular matrix}%
   {RC singular matrix}%
\Index{texVectorSpace}
   {\rcd vector space}%
   {rcd vector space}%
\Index{texCartesian}
   {reduced Cartesian product of bundles}%
   {reduced Cartesian product of bundles}%
\Index{texCartesian}
   {reduced Cartesian product of total spaces}%
   {reduced Cartesian product of total spaces}%
\Index{texBundleRelation,texBundleRelation}
   {reduced fibered correspondence from $\mathcal{A}$ to $\mathcal{B}$}%
   {reduced fibered correspondence from A to B}%
\Index{texBundleRelation}
   {reduced fibered correspondence in $\mathcal{A}$}%
   {reduced fibered correspondence in A}%
\Index{texBiring}
   {reducible biring}%
   {reducible biring}%
\Index{texRefernceFrame}
   {reference frame in event space}%
   {reference frame in event space}%
\Index{texDrcReferenceFrame}
   {reference frame manifold}%
   {reference frame manifold}%
\Index{texBundleRelation}
   {reflexive $2$\Hyph ary fibered relation}%
   {reflexive 2 ary fibered relation}%
\Index{texTstarRepresentation}
   {representation of group}%
   {}%
\SubIndex{texTstarRepresentation}
   {contravariant \Ts}%
   {Tstar contravariant representation of group}%
\SubIndex{texTstarRepresentation}
   {covariant \Ts}%
   {Tstar covariant representation of group}%
\SubIndex{texDrcBasis}
   {\drc linear \sT}%
   {linear representation of group}%
\SubIndex{texTstarRepresentation}
   {effective}%
   {effective representation of group}%
\SubIndex{texDrcBasis}
   {\rcd}%
   {rcd linear representation of group}%
\SubIndex{texTstarRepresentation}
   {\sT}%
   {starT representation of group}%
\SubIndex{texTstarRepresentation}
   {\Ts}%
   {Tstar representation of group}%
\Index{texRepresentation}
   {representation of group}%
   {representation of group}%
\Index{texDrcBasis}
   {representative of geometrical object in vector space}%
   {representative of geometrical object, vector space}%
\Index{texBasis}
   {representative of geometrical object in vector space}%
   {representative of geometrical object, vector space}%
\Index{texBundleRelation}
   {restriction of correspondence $\Phi$ to set $C$}%
   {restriction of correspondence}%
\Index{texLie}
   {right invariant vector field}%
   {right invariant vector}%
\Index{texVectorSpace}
   {right module}%
   {right module}%
\Index{texTstarRepresentation}
   {right shift}%
   {right shift}%
\Index{texRepresentation}
   {right shift on group}%
   {right shift, group}%
\Index{texLie}
   {right structural constant of Lie algebra}%
   {right structural constant of Lie algebra}%
\Index{texVectorSpace}
   {right vector space}%
   {right vector space}%
\Index{texRepresentation}
   {right-side contravariant representation of group}%
   {right-side contravariant representation}%
\Index{texRepresentation}
   {right-side covariant representation of group}%
   {right-side covariant representation}%
\Index{texTstarMorphism}
   {right-side representation of $\mathfrak{F}$\Hyph algebra $A$ in set $M$}%
   {right-side representation of algebra}%
\Index{texFiberedAlgebra}
   {right-side representation of fibered $\mathfrak{F}$\Hyph algebra}%
   {right-side representation of fibered F-algebra}%
\Index{texRepresentation}
   {right-side representation of group}%
   {right-side representation of group}%
\Index{texTstarMorphism}
   {right-side transformation}%
   {right-side transformation}%
\Index{texRepresentation}
   {right-side transformation}%
   {right-side transformation}%
\Index{texRepresentation}
   {row vector}%
   {row vector}%
\Index{texVectorSpace}
   {$R\star$\Hyph module}%
   {Rstar-module}%
\SetIndexSpace%
\Index{texNewton}
   {scalar potential}%
   {scalar potential}%
\Index{texNewton}
   {second Newton law}%
   {Second Newton law}%
\Index{texTstarMorphism}
   {single transitive representation of algebra $A$}%
   {single transitive representation of algebra}%
\Index{texFiberedAlgebra}
   {single transitive representation of fibered $\mathfrak{F}$\Hyph algebra}%
   {single transitive representation of fibered F-algebra}%
\Index{texRepresentation}
   {single transitive representation of group}%
   {single transitive representation of group}%
\Index{texTstarRepresentation}
   {space of orbits of \Ts representation}%
   {space of orbits of Ts representation}%
\Index{texTidal}
   {speed of deviation}%
   {speed of deviation}%
\Index{texDrcMorphism}
   {$(S\RCstar,T\RCstar)$\Hyph linear map of vector spaces}%
   {src trc linear map of vector spaces}%
\Index{texTstarRepresentation}
   {stability group}%
   {stability group}%
\Index{texDrcBasis}
   {standard coordinates of basis}%
   {standard coordinates of basis}%
\Index{texBasis}
   {standard coordinates of basis}%
   {standard coordinates of basis}%
\Index{texBiring}
   {standard representation of matrix}%
   {Standard representation}%
\Index{texVectorSpace}
   {$\star D$\hyph vector space}%
   {starD-vector space}%
\Index{texLinearMap}
   {$\star D$\Hyph product of \drc linear map $A$ over scalar}%
   {starD product of drc linear map over scalar}%
\Index{texVectorSpace}
   {$\star R$\hyph module}%
   {starR-module}%
\Index{texTstarRepresentation}
   {\sT shift}%
   {starT shift}%
\Index{texFiberedGroup}
   {\sT shift on fibered group}%
   {starT shift, fibered group}%
\Index{texTstarMorphism}
   {\sT representation of $\mathfrak{F}$\Hyph algebra $A$ in set $M$}%
   {starT representation of algebra}%
\Index{texFiberedAlgebra}
   {\sT representation of fibered $\mathfrak{F}$\Hyph algebra}%
   {starT representation of fibered F-algebra}%
\Index{texFiberedGroup}
   {\sT representation of fibered group}%
   {starT representation of fibered group}%
\Index{texTstarMorphism}
   {\sT transformation}%
   {starT transformation}%
\Index{texFiberedAlgebra}
   {\sT transformation on bundle}%
   {starT transformation of bundle}%
\SubIndex{}
   {nonsingular}%
   {nonsingular transformation of bundle}%
\Index{texBundle}
   {subbundle}%
   {subbundle}%
\Index{texLinearMap}
   {sum of \drc linear maps}%
   {sum of drc linear maps, drc vector spaces}%
\Index{texDrcBasis}
   {sum of geometrical objects}%
   {sum of geometrical objects}%
\Index{texBasis}
   {sum of geometrical objects in vector space}%
   {sum of geometrical objects, vector space}%
\Index{texBundleRelation}
   {symmetric $2$\Hyph ary fibered relation}%
   {symmetric 2 ary fibered relation}%
\Index{texBasis}
   {symmetry group}%
   {symmetry group}%
\Index{texDrcBasis}
   {symmetry group}%
   {SymmetryGroup}%
\Index{texGenRelativity}
   {synchronization of reference frame}%
   {synchronization of reference frame}%
\SetIndexSpace%
\Index{texLie}
   {tensor product of representations}%
   {tensor product of representations}%
\Index{texCalculus}
   {topological \drc vector space}%
   {topological drc vector space}%
\Index{texCalculus}
   {topological skew field}%
   {topological skew field}%
\Index{texGeomObject}
   {torsion form}%
   {torsion form}%
\Index{texGeomObject}
   {torsion tensor}%
   {torsion tensor}%
\Index{texFiberedMorphism}
   {tower of bundles}%
   {tower of bundles}%
\Index{texTstarMorphism}
   {transformation of set}%
   {transformation of set}%
\Index{texDrcBasis}
   {transformation on basis manifold}%
   {}%
\SubIndex{texDrcBasis}
   {active}%
   {active transformation, vector space}%
\SubIndex{texTypeBasis}
   {affine}%
   {affine transformation}%
\SubIndex{texTypeBasis}
   {movement}%
   {movement transformation}%
\SubIndex{texDrcBasis}
   {passive}%
   {passive transformation, vector space}%
\SubIndex{texTypeBasis}
   {quasi affine}%
   {quasi affine transformation}%
\SubIndex{texTypeBasis}
   {quasi movement}%
   {quasi movement}%
\Index{texFiberedAlgebra}
   {transformation on bundle}%
   {transformation of bundle}%
\Index{texBundleRelation}
   {transitive $2$\Hyph ary fibered relation}%
   {transitive 2 ary fibered relation}%
\Index{texTstarMorphism}
   {transitive representation of $\mathfrak{F}$\Hyph algebra $A$}%
   {transitive representation of algebra}%
\Index{texFiberedAlgebra}
   {transitive representation of fibered $\mathfrak{F}$\Hyph algebra}%
   {transitive representation of fibered F-algebra}%
\Index{texRepresentation}
   {transitive representation of group}%
   {transitive representation of group}%
\Index{texVectorSpace}
   {\Ts linear composition of  vectors}%
   {linear composition of  vectors}%
\Index{texVectorSpace}
   {\Ts matrices vector space}%
   {matrices vector space}%
\Index{texTstarRepresentation}
   {\Ts shift}%
   {Tstar shift}%
\Index{texTstarMorphism}
   {\Ts representation of $\mathfrak{F}$\Hyph algebra $A$ in set $M$}%
   {Tstar representation of algebra}%
\Index{texFiberedAlgebra}
   {\Ts representation of fibered $\mathfrak{F}$\Hyph algebra}%
   {Tstar representation of fibered F-algebra}%
\Index{texFiberedGroup}
   {\Ts representation of fibered group}%
   {Tstar representation of fibered group}%
\Index{texTstarMorphism}
   {\Ts transformation}%
   {Tstar transformation}%
\Index{texFiberedAlgebra}
   {\Ts transformation on bundle}%
   {Tstar transformation of bundle}%
\Index{texFiberedGroup}
   {twin representations of fibered group}%
   {twin representations of fibered group}%
\Index{texTstarRepresentation}
   {twin representations of group}%
   {twin representations of group}%
\Index{texLinearMap}
   {twin representations of skew field}%
   {twin representations of skew field}%
\SetIndexSpace%
\Index{texVectorSpace}
   {unitarity law}%
   {}%
\SubIndex{texVectorSpace}
   {for \Ts vector space}%
   {unitarity law, Tstar vector space}%
\SetIndexSpace%
\Index{texPolymodule}
   {($D_1\RCstar$, ..., $D_n\RCstar$)\hyph vector space}%
   {(d1rc,dnrc)-vector space}%
\Index{texPolymodule}
   {($S\star$, $\star T$)\hyph vector space}%
   {(Sstar,starT)-vector space}%
\Index{texCalculus}
   {valued skew field}%
   {valued skew field}%
\Index{texFiberedAlgebra}
   {vector bundle}%
   {vector bundle}%
\Index{texNewton}
   {vector potential}%
   {vector potential}%
\Index{texVectorSpace}
   {vector space type}%
   {vector space type}%

\CloseIndex

\def\indexname{Special Symbols and Notations}
\OpenIndex

\SetIndexSpace%
\Symb{texBiring}
   {$(^{\gi a}_{\gi b})$\hyph\CR quasideterminant}%
   {a b CR quasideterminant definition}%
\Symb{texBiring}
   {$(^{\gi a}_{\gi b})$\hyph \RC quasideterminant}%
   {a b RC-quasideterminant definition}%
\Symb{texBiring}
   {minor}%
   {A from b a}%
\Symb{texBiring}
   {minor}%
   {A from columns T}%
\Symb{texBiring}
   {minor}%
   {A from rows S}%
\Symb{texBiring}
   {minor}%
   {A without column a}%
\Symb{texBiring}
   {minor}%
   {A without columns T}%
\Symb{texBiring}
   {minor}%
   {A without row b}%
\Symb{texBiring}
   {minor}%
   {A without rows S}%
\Symb{texPolymodule}
   {active transformation}%
   {active transformation}%
\Symb{texTypeBasis}
   {affine space}%
   {affine space}%
\Symb{texBasis}
   {affine space}%
   {An}%
\Symb{texBiring}
   {\subs row ($c$\hyph row) of matrix}%
   {c row}%
\Symb{texBiring}
   {\CR power of element $A$ of biring}%
   {cr power}%
\Symb{texBiring}
   {\CR inverse element of biring}%
   {cr-inverse element}%
\Symb{texBiring}
   {\CR product of matrices}%
   {cr-product of matrices}%
\Symb{texVectorSpace}
   {\dcr vector}%
   {dcr vector}%
\Symb{texLie}
   {derivative of left shift}%
   {derivative of left shift}%
\Symb{texLie}
   {derivative of left shift}%
   {derivative of left shift, 1-Parameter Group}%
\Symb{texLie}
   {derivative of right shift}%
   {derivative of right shift}%
\Symb{texLie}
   {}%
   {derivative of right shift}%
\Symb{texLie}
   {derivative of right shift}%
   {derivative of right shift, 1-Parameter Group}%
\Symb{texLie}
   {derivative of left shift}%
   {derivative of Tstar shift}%
\Symb{texVectorSpace}
   {\drc vector}%
   {drc vector}%
\Symb{texAffine}
   {derivative}%
   {overline nabla_l, definition 2}%
\Symb{texPolymodule}
   {passive transformation}%
   {passive transformation}%
\Symb{texBiring}
   {\sups row ($r$\hyph row) of matrix}%
   {r row}%
\Symb{texBiring}
   {\RC power of element $A$ of biring}%
   {rc power}%
\Symb{texBiring}
   {\RC inverse element of biring}%
   {rc-inverse element}%
\Symb{texBiring}
   {\RC product of matrices}%
   {rc-product of matrices}%
\Symb{texBiring}
   {\RC quasideterminant}%
   {RC-quasideterminant definition}%
\Symb{texTstarRepresentation}
   {right shift}%
   {right shift}%
\Symb{texFiberedGroup}
   {\sT shift}%
   {starT shift, fibered group}%
\Symb{texTstarRepresentation}
   {left shift}%
   {Tstar shift}%
\Symb{texFiberedGroup}
   {\Ts shift}%
   {Tstar shift, fibered group}%
\Symb{texRefernceFrame}
   {anholonomic coordinates of vector}%
   {vector anholonomic coordinates}%
\Symb{texRefernceFrame}
   {holonomic coordinates of vector}%
   {vector holonomic coordinates}%

\SetIndexSpace%
\Symb{texBasis}
   {basis manifold of affine space}%
   {BAn}%
\Symb{texBasis}
   {basis manifold of vector space $\mathcal{V}$}%
   {basis manifold of vector space}%
\Symb{texBasis}
   {basis manifold of vector space}%
   {basis manifold of vector space}%
\Symb{texBasis}
   {basis manifold of central affine space}%
   {BCAn}%
\Symb{texBasis}
   {basis manifold of Euclid space}%
   {BEn}%
\Symb{texBundle}
   {Cartesian power $\mathcal{A}$ of bundle $\mathcal{B}$}%
   {Cartesian power of bundle}%
\Symb{texBundle}
   {Cartesian power $A$ of set $B$}%
   {Cartesian power of set}%
\Symb{texTypeBasis}
   {basis manifold of affine space}%
   {FAn}%
\Symb{texTypeBasis}
   {basis manifold of central affine space}%
   {FCAn}%
\Symb{texTypeBasis}
   {basis manifold of Euclid space}%
   {FEn}%

\SetIndexSpace%
\Symb{texBasis}
   {central affine space}%
   {CAn}%
\Symb{texTypeBasis}
   {central affine space}%
   {central affine space}%
\Symb{texLie}
   {left structural constant of Lie algebra}%
   {left structural constant of Lie algebra}%
\Symb{texLie}
   {right structural constant of Lie algebra}%
   {right structural constant of Lie algebra}%

\SetIndexSpace%
\Symb{texLieRepresentation}
   {basis vector of \sT representation}%
   {basis vector of starT representation}%
\Symb{texLieRepresentation}
   {basis vector of \sT representation}%
   {basis vector of starT representation, coordinates}%
\Symb{texLieRepresentation}
   {basis vector of \Ts representation}%
   {basis vector of Tstar representation}%
\Symb{texLieRepresentation}
   {basis vector of \Ts representation}%
   {basis vector of Tstar representation, coordinates}%
\Symb{texVectorSpace}
   {\subs rows \dcr vector space}%
   {c rows dcr vector space}%
\Symb{texCalculus}
   {differential of function}%
   {differential, drc vector space to drc vector space}%
\Symb{texCalculus}
   {differential of function}%
   {differential, drc vector space to skew field}%
\Symb{texVectorSpace}
   {\drc coordinate vector space}%
   {drc coordinate vector space}%
\Symb{texVectorSpace}
   {matrices vector space}%
   {matrices vector space}%
\Symb{texAffine}
   {Cartan derivative}%
   {overbrace D}%
\Symb{texAffine}
   {derivative}%
   {overline D}%
\Symb{texCalculus}
   {partial derivative of mapping $\Vector f$ with respect to variable $v^{\gi a}$}%
   {partial derivative of mapping, 1, drc vector space}%
\Symb{texCalculus}
   {partial derivative of mapping $f$ with respect to variable $v^{\gi a}$}%
   {partial derivative of mapping, 1, skew field}%
\Symb{texVectorSpace}
   {\sups rows \drc vector space}%
   {r rows drc vector space}%
\Symb{texTidal}
   {speed of deviation}%
   {speed of deviation}%
\Symb{texVectorSpace}
   {vector space type}%
   {vector space type}%

\SetIndexSpace%
\Symb{texTypeBasis}
   {affine basis}%
   {Affine Basis}%
\Symb{texBasis}
   {affine basis}%
   {Affine Basis}%
\Symb{texTypeBasis}
   {basis}%
   {basis}%
\Symb{texBasis}
   {basis of vector space}%
   {Basis e}%
\Symb{texBasis}
   {basis in vector space $\mathcal{V}$}%
   {basis in V}%
\Symb{texVectorSpace}
   {basis of vector space}%
   {basis, vector space}%
\Symb{texPolymodule}
   {basis of $(n)$\hyph vector space}%
   {basis,n vector space}%
\Symb{texCartesian}
   {Cartesian power of total spaces}%
   {Cartesian power of total spaces}%
\Symb{texCartesian}
   {Cartesian product of total spaces}%
   {Cartesian product of total spaces, definition 1}%
\Symb{texBasis}
   {central affine basis}%
   {Central Affine Basis}%
\Symb{texRefernceFrame}
   {form of reference frame}%
   {dual forms, reference frame}%
\Symb{texBasis}
   {Euclid space}%
   {En}%
\Symb{texTypeBasis}
   {Euclid space}%
   {En}%
\Symb{texTypeBasis}
   {pseudo Euclid space}%
   {Enm}%
\Symb{texBasis}
   {pseudo Euclid space}%
   {Enm}%
\Symb{texFiberedAlgebra}
   {identical transformation of bundle}%
   {identical transformation of bundle}%
\Symb{texBasis}
   {orthonornal basis}%
   {Orthonornal Basis}%
\Symb{texCartesian}
   {reduced Cartesian product of total spaces}%
   {reduced Cartesian product of total spaces, definition 1}%
\Symb{texFiberedAlgebra}
   {set of nonsingular \sT transformations of bundle $\mathcal{E}$}%
   {set of starT nonsingular transformations of bundle}%
\Symb{texFiberedAlgebra}
   {set of nonsingular \Ts transformations of bundle $\mathcal{E}$}%
   {set of Tstar nonsingular transformations of bundle}%
\Symb{texBasis}
   {standard coordinates of basis}%
   {standard coordinates of basis}%
\Symb{texRefernceFrame}
   {standard coordinates of reference frame}%
   {standard coordinates of reference frame}%
\Symb{texRefernceFrame}
   {vector field of reference frame}%
   {vector field of reference frame}%
\Symb{texBasis}
   {vector of basis}%
   {vector of basis}%

\SetIndexSpace%
\Symb{texVectorSpace}
   {coordinates of basis in \subs rows \dcr vector space}%
   {basis coordinates, c rows dcr vector space}%
\Symb{texVectorSpace}
   {coordinates of basis in \sups rows \drc vector space}%
   {basis coordinates, r rows drc vector space}%
\Symb{texVectorSpace}
   {basis for \subs rows \dcr vector space}%
   {basis, c rows dcr vector space}%
\Symb{texVectorSpace}
   {basis for \sups rows \drc vector space}%
   {basis, r rows drc vector space}%
\Symb{texDiffEq}
   {central affine basis}%
   {Central Affine Basis}%
\Symb{texBundle}
   {fibered morphism from bundle $\mathcal{A}$ into $\mathcal{B}$}%
   {fibered morphism from A into B}%
\Symb{texBundleRelation}
   {filter $\mathfrak{F}$ converges to set $A$}%
   {filter converges}%
\Symb{texFiberedAlgebra}
   {homomorphism of fibered $\mathfrak{F}$\Hyph algebras}%
   {homomorphism of fibered F-algebras}%
\Symb{texBundleRelation}
   {inverse fibered correspondence}%
   {inverse fibered correspondence, 1}%
\Symb{texBundleRelation}
   {inverse reduced fibered correspondence}%
   {inverse reduced fibered correspondence, 1}%
\Symb{texCartesian}
   {map to Cartesian product}%
   {map to Cartesian product}%
\Symb{texTstarRepresentation}
   {representation orbit of group $G$}%
   {orbit of representation of group}%
\Symb{texTypeBasis}
   {orthonornal basis}%
   {Orthonornal Basis}%
\Symb{texRefernceFrame}
   {reference frame}%
   {reference frame}%
\Symb{texRefernceFrame}
   {reference frame, extensive definition}%
   {reference frame, extensive definition}%
\Symb{texPolymodule}
   {standard coordinates of basis}%
   {standard coordinates of basis}%
\Symb{texPolymodule}
   {vector of basis}%
   {vector of basis}%

\SetIndexSpace%
\Symb{texVectorSpace}
   {\CR matrix group}%
   {cr-matrix group}%
\Symb{texFiberedMorphism}
   {fibered little group of section $h$}%
   {fibered little group}%
\Symb{texFiberedMorphism}
   {fibered stability group of section $h$}%
   {fibered stability group}%
\Symb{texLie}
   {algebra Lie of group Lie}%
   {g}%
\Symb{texLie}
   {left defined algebra Lie of group Lie}%
   {gl}%
\Symb{texTypeBasis}
   {affine transformation group}%
   {GLAn}%
\Symb{texBasis}
   {affine transformation group}%
   {GLAn}%
\Symb{texLie}
   {right defined algebra Lie of group Lie}%
   {gr}%
\Symb{texBasis}
   {group of homomorphisms of vector space $\mathcal{V}$}%
   {GV}%
\Symb{texTstarRepresentation}
   {little group of $x$}%
   {little group}%
\Symb{texFiberedGroup}
   {orbit of effective covariant \sT representation of fibered group}%
   {orbit of effective starT covariant representation of fibered group}%
\Symb{texTstarRepresentation}
   {orbit of effective covariant \sT representation of group}%
   {orbit of effective starT covariant representation of group}%
\Symb{texFiberedGroup}
   {orbit of effective covariant \Ts representation of fibered group}%
   {orbit of effective Tstar covariant representation of fibered group}%
\Symb{texTstarRepresentation}
   {orbit of effective covariant \Ts representation of group}%
   {orbit of effective Tstar covariant representation of group}%
\Symb{texVectorSpace}
   {\RC matrix group}%
   {rc-matrix group}%
\Symb{texTstarRepresentation}
   {stability group of $x$}%
   {stability group}%

\SetIndexSpace%
\Symb{texBiring}
   {Hadamard inverse of matrix}%
   {Hadamard inverse of matrix}%
\Symb{texLinearMap}
   {\rcd vector space of \drc linear maps}%
   {rcd vector space of drc linear maps}%

\SetIndexSpace%
\Symb{texLieRepresentation}
   {infinitesimal generator of representation}%
   {infinitesimal generator of representation}%
\Symb{texLinearLie}
   {Lie group infinitesimal generators}%
   {Lie group infinitesimal generators}%

\SetIndexSpace%
\Symb{texRepresentation}
   {left shift}%
   {left shift}%
\Symb{texDiffProperty}
   {Lie derivative of connection}%
   {Lie derivative of connection}%
\Symb{texDiffProperty}
   {Lie derivative of metric}%
   {Lie derivative of metric}%
\Symb{texBundleRelation}
   {limit of correspondence $\Phi$ with respect to the filter $\mathfrak{F}$}%
   {limit of correspondence with respect to the filter}%
\Symb{texBasis}
   {passive transformation}%
   {passive transformation}%
\Symb{texRepresentation}
   {set of left-side nonsingular transformations of set $M$}%
   {set of left-side nonsingular transformations}%

\SetIndexSpace%
\Symb{texTstarMorphism}
   {set of \sT transformations of set $M$}%
   {set of starT transformations}%
\Symb{texTstarMorphism}
   {set of \Ts transformations of set $M$}%
   {set of Tstar transformations}%
\Symb{texTstarRepresentation}
   {space of orbits of effective \sT covariant representation of the group}%
   {space of orbits of effective sT representation}%
\Symb{texTstarRepresentation}
   {space of orbits of effective \Ts covariant representation of the group}%
   {space of orbits of effective Ts representation}%
\Symb{texTstarRepresentation}
   {space of orbits of \Ts representation $f$ of group $G$ in set $M$}%
   {space of orbits of Ts representation}%

\SetIndexSpace%
\Symb{texBasis}
   {geometrical object in coordinate representation}%
   {geometrical object, coordinate vector space}%
\Symb{texBasis}
   {geometrical object}%
   {geometrical object, vector space}%
\Symb{texFiberedGroup}
   {orbit of representation of fibered group $\mathcal{G}$}%
   {orbit of representation of fibered group}%
\Symb{texRepresentation}
   {orbit of representation of the group $G$}%
   {orbit of representation of group}%

\SetIndexSpace%
\Symb{texBundle}
   {bundle}%
   {bundle}%
\Symb{texFiberedMorphism}
   {bundle of level $2$}%
   {bundle of level 2}%
\Symb{texFiberedMorphism}
   {bundle of level $n$}%
   {bundle of level n}%
\Symb{texCartesian}
   {Cartesian power of bundle}%
   {Cartesian power of bundle}%
\Symb{texCartesian}
   {Cartesian product of bundles}%
   {Cartesian product of bundles, definition 1}%
\Symb{texCartesian}
   {reduced Cartesian product of bundles}%
   {reduced Cartesian product of bundles, definition 1}%
\Symb{texFiberedAlgebra}
   {set of nonsingular \sT transformations of bundle $\bundle{}pE{}$}%
   {set of starT nonsingular transformations of bundle, projection}%
\Symb{texFiberedAlgebra}
   {set of nonsingular \Ts transformations of bundle $\bundle{}pE{}$}%
   {set of Tstar nonsingular transformations of bundle, projection}%

\SetIndexSpace%
\Symb{texBasis}
   {active transformation}%
   {active transformation}%
\Symb{texAffine}
   {Cartan curvature}%
   {Cartan curvature}%
\Symb{texVectorSpace}
   {\CR rank of matrix}%
   {cr-rank of matrix}%
\Symb{texBundleRelation}
   {diagonal in bundle  $\bundle{}pA{}$}%
   {diagonal in bundle, 2}%
\Symb{texBundleRelation}
   {diagonal in bundle $\mathcal{A}$}%
   {diagonal in reduced bundle, 2}%
\Symb{texAffine}
   {curvature}%
   {GLn curvature_overline}%
\Symb{texVectorSpace}
   {\RC rank of matrix}%
   {rc-rank of matrix}%
\Symb{texRepresentation}
   {right shift}%
   {right shift}%
\Symb{texRepresentation}
   {set of right-side nonsingular transformations of set $M$}%
   {set of right-side nonsingular transformations}%

\SetIndexSpace%
\Symb{texBundleRelation}
   {composition of fibered correspondences}%
   {composition of fibered correspondences}%
\Symb{texBundleRelation}
   {inverse fibered correspondence}%
   {inverse fibered correspondence, 2}%
\Symb{texBundleRelation}
   {inverse reduced fibered correspondence}%
   {inverse reduced fibered correspondence, 2}%
\Symb{texVectorSpace}
   {linear span in vector space}%
   {linear span, vector space}%

\SetIndexSpace%
\Symb{texLie}
   {tangent plane to group $G$}%
   {TaG}%

\SetIndexSpace%
\Symb{texBasis}
   {coordinate vector space}%
   {coordinate vector space}%
\Symb{texBasis}
   {coordinates in vector space}%
   {coordinates in vector space}%
\Symb{texVectorSpace}
   {\dcr vector space}%
   {left CR vector space}%
\Symb{texVectorSpace}
   {\drc vector space}%
   {left RC vector space}%
\Symb{texLinearMap}
   {($S$, $T$)\hyph bimodule}%
   {R S bimodule}%
\Symb{texVectorSpace}
   {\crd vector space}%
   {right CR vector space}%
\Symb{texVectorSpace}
   {\rcd vector space}%
   {right RC vector space}%
\Symb{texBasis}
   {vector space}%
   {V}%

\SetIndexSpace%
\Symb{texPolymodule}
   {geometrical object in coordinate representation		defined in vector space}%
   {geometrical object, coordinate vector space}%
\Symb{texPolymodule}
   {geometrical object in vector space}%
   {geometrical object, vector space}%

\SetIndexSpace%
\Symb{texRefernceFrame}
   {anholonomic coordinate}%
   {x(k)}%

\SetIndexSpace%
\Symb{texBundleRelation}
   {diagonal in bundle $\mathcal{A}$}%
   {diagonal in bundle, 1}%

\SetIndexSpace%
\Symb{texTidal}
   {deviation of trajectories}%
   {deviation of trajectories}%
\Symb{texRepresentation}
   {identical transformation}%
   {identical transformation}%
\Symb{texTstarMorphism}
   {identical transformation}%
   {identical transformation}%
\Symb{texBasis}
   {image of vector $\Vector e_k\in\Basis e$ under isomorphism to coordinate vector space}%
   {image of vector e_k, coordinate vector space}%
\Symb{texBiring}
   {Kronecker symbol}%
   {Kronecker symbol}%

\SetIndexSpace%
\Symb{texRefernceFrame}
   {anholonomic coordinates of connection}%
   {anholonomic coordinates of connection}%
\Symb{texAffine}
   {Cartan symbol}%
   {Cartan symbol}%
\Symb{texAffine}
   {connection}%
   {conection overline}%
\Symb{texRefernceFrame}
   {holonomic coordinates of connection}%
   {holonomic coordinates of connection}%
\Symb{texAffine}
   {Cartan connection}%
   {overbrace Gamma i kl}%
\Symb{texBundle}
   {set of sections of bundle}%
   {set of sections of bundle}%

\SetIndexSpace%
\Symb{texLie}
   {inverse operator to operator $\psi_l$}%
   {inverse operator to operator psi l}%
\Symb{texLie}
   {inverse operator to operator $\psi_r$}%
   {inverse operator to operator psi r}%

\SetIndexSpace%
\Symb{texRefernceFrame}
   {anholonomity object}%
   {anholonomity object}%

\SetIndexSpace%
\Symb{texLie}
   {basic operator of group Lie}%
   {Lie Basic Operator L}%
\Symb{texLie}
   {}%
   {Lie Basic Operator L}%
\Symb{texLie}
   {basic operator of group Lie}%
   {Lie Basic Operator L, 1-Parameter Group}%
\Symb{texLie}
   {basic operator of group Lie}%
   {Lie Basic Operator R}%
\Symb{texLie}
   {}%
   {Lie Basic Operator R}%
\Symb{texLie}
   {basic operator of group Lie}%
   {Lie Basic Operator R, 1-Parameter Group}%

\SetIndexSpace%
\Symb{texRefernceFrame}
   {coordinate reference frame}%
   {coordinate reference frame, extensive definition}%
\Symb{texCalculus}
   {partial derivative of mapping $\Vector f$ with respect to variable $v^{\gi a}$}%
   {partial derivative of mapping, 2, drc vector space}%
\Symb{texCalculus}
   {partial derivative of mapping $f$ with respect to variable $v^{\gi a}$}%
   {partial derivative of mapping, 2, skew field}%
\Symb{texRefernceFrame}
   {derivative $e_{(k)}$}%
   {partial(k)}%

\SetIndexSpace%
\Symb{texLie}
   {Lie group composition law}%
   {Lie group composition law}%

\SetIndexSpace%
\Symb{texAffine}
   {}%
   {overbrace nabla_l}%
\Symb{texAffine}
   {Cartan derivative}%
   {overbrace nabla_l}%
\Symb{texAffine}
   {derivative}%
   {overline nabla_l, definition 1}%

\SetIndexSpace%
\Symb{texBundleRelation}
   {restriction of correspondence $\Phi$ to set $C$}%
   {restriction of correspondence}%

\SetIndexSpace%
\Symb{texCartesian}
   {Cartesian product of bundles}%
   {Cartesian product of bundles, definition 2}%
\Symb{texCartesian}
   {Cartesian product of total spaces}%
   {Cartesian product of total spaces, definition 2}%
\Symb{texCartesian}
   {reduced Cartesian product of bundles}%
   {reduced Cartesian product of bundles, definition 2}%
\Symb{texCartesian}
   {reduced Cartesian product of total spaces}%
   {reduced Cartesian product of total spaces, definition 2}%

\SetIndexSpace%
\Symb{texBundle}
   {fibered subset}%
   {fibered subset}%
\Symb{texBundle}
   {subbundle}%
   {subbundle}%

\CloseIndex

\end{document}